\documentclass{aa}  
\usepackage{ctable}
\usepackage{natbib,twoopt}
\usepackage{graphicx}
\usepackage{subcaption}
\usepackage{cleveref}
\usepackage{amsmath,amssymb}
\usepackage{threeparttable}

\bibpunct{(}{)}{;}{a}{}{,} %% natbib format for A&A and ApJ
\makeatletter
\newcommandtwoopt{\citeads}[3][][]{\href{http://adsabs.harvard.edu/abs/#3}%
{\def\hyper@linkstart##1##2{}%
\let\hyper@linkend\@empty\citealp[#1][#2]{#3}}}
\newcommandtwoopt{\citepads}[3][][]{\href{http://adsabs.harvard.edu/abs/#3}%
{\def\hyper@linkstart##1##2{}%
\let\hyper@linkend\@empty\citep[#1][#2]{#3}}}
\newcommandtwoopt{\citetads}[3][][]{\href{http://adsabs.harvard.edu/abs/#3}%
{\def\hyper@linkstart##1##2{}%
\let\hyper@linkend\@empty\citet[#1][#2]{#3}}}
\newcommandtwoopt{\citeyearads}[3][][]%
{\href{http://adsabs.harvard.edu/abs/#3}
{\def\hyper@linkstart##1##2{}%
\let\hyper@linkend\@empty\citeyear[#1][#2]{#3}}}

\def\nlte{non-LTE}

\def\nlte{non-LTE}

\def\Msun{\ensuremath{\,M_\odot\,}}
\def\Lsun{\ensuremath{\,L_\odot\,}}
\def\Rsun{\ensuremath{\,R_\odot\,}}

\def\Mdot{\dot{M}}

\makeatother

\usepackage{txfonts}
\begin{document}

\title{Subsonic structure and optically thick winds \\from Wolf--Rayet stars}

   \author{L. Grassitelli
          \inst{1}\fnmsep\thanks{e-mail: luca@astro.uni-bonn.de}
        ,          
        N. Langer\inst{1}
        ,
                 N.J. Grin\inst{1}
        ,
                        J. Mackey\inst{2}  
                 ,
        J.M. Bestenlehner\inst{3,4}       
         and
                G. Gr\"afener\inst{1}
}

   \institute{ Argelander-Institut f\"ur Astronomie, Universit\"at Bonn, Auf
              dem H\"ugel 71, 53121 Bonn, Germany
        \and
 Dublin Institute for Advanced Studies, Dunsink Observatory, Dunsink Lane, Castleknock, Dublin 15, Ireland
        \and
        Department of Physics and Astronomy, University of Sheffield, Hicks Building, Hounsfield Rd, Sheffield, S3 7RH, UK
        \and
Max-Planck-Institut f\"ur Astronomie, K\"onigstuhl 17. 69117 Heidelberg, German
         }

%  \date{Received February 29, 2012; accepted --}
   \date{Received //, 2017}

% \abstract{}{}{}{}{} 
% 5 {} token are mandatory
 \abstract{
 Mass loss by stellar wind is a key agent in the evolution and spectroscopic appearance of massive main sequence and post-main sequence stars. In Wolf--Rayet stars the winds can be so dense and so optically thick that the photosphere appears in the highly supersonic part of the outflow, veiling the underlying subsonic part of the star, and leaving the initial acceleration of the wind inaccessible to observations. Here we  investigate the conditions and the structure of the subsonic part of the outflow of Galactic Wolf--Rayet stars, in particular of the WNE  subclass; our focus is on the conditions at the sonic point of their winds. 

 We compute 1D hydrodynamic stellar structure models for massive helium stars adopting outer boundaries at the sonic point. We find that the outflows of our models are accelerated to supersonic velocities by the radiative force from opacity bumps either at temperatures of the order of 200\,kK by the  iron opacity bump or of the order of 50\,kK by the helium-II opacity bump. For a given mass-loss rate, the diffusion approximation for radiative energy transport allows us to define the temperature gradient based purely on the local thermodynamic conditions. 
For a given mass-loss rate, this implies that the conditions in the subsonic part of the outflow are independent from the detailed physical conditions in the supersonic part. Stellar atmosphere calculations can therefore adopt our hydrodynamic models as ab initio input for the subsonic structure.\\
The close proximity to the Eddington limit at the sonic point allows us to construct a {\it Sonic HR diagram}, relating the sonic point temperature to the luminosity-to-mass ratio  and the stellar mass-loss rate, thereby  constraining the sonic point conditions, the subsonic structure, and the stellar wind mass-loss rates of WNE stars from observations. 

 The minimum stellar wind mass-loss rate necessary to have the flow accelerated to supersonic velocities by the iron opacity bump is derived. A comparison of the observed parameters of Galactic WNE stars to this minimum mass-loss rate indicates that these stars have their winds launched to supersonic velocities by the radiation pressure arising from the iron opacity bump.        
Conversely, stellar models which do not show transonic flows from the iron opacity bump form low-density  extended envelopes. We derive an analytic criterion for the appearance of envelope inflation and of a density inversion in the outer sub-photospheric layers.   
 }


      \authorrunning{Grassitelli et al.2017}
         
   \maketitle

%       \titlerunning
%
%________________________________________________________________

\section{Introduction}

A star loses energy from its surface not only in the form of radiation, but also in the form of a stream of particles, the stellar wind \citep{1958Parker}. Stellar winds are outflows of material from a star which play a major role especially in massive star evolution. Stellar winds can in fact lead massive stars to lose a significant fraction of their initial mass, thus enriching the interstellar medium and  influencing the structure of the outer layers  and  the spectroscopic appearance of these stars \citep{1986Chiosi,1991Lamers,2012Langer}. 

The proximity of massive stars to the Eddington limit in the upper part of the Hertszprung--Russell (HR) diagram can lead to dense outflows. High mass-loss rates can veil the hydrostatic layers and result in the appearance of emission lines in the spectra of luminous stars  \citep[e.g.][]{2006Hamann,2008Grafener,2014Bestenlehner}. In this case part of the radiation is scattered back from the radially expanding atmosphere, giving feedback in the form of a back-warming of the outer hydrostatic layers. This feedback is often described as proportional to the frequency independent mean optical depth $\tau$, which is the average number of photon mean-free paths along the line of sight \citep{1978Mihalas}. For optically thin winds the photosphere arises in the subsonic part of the outflow. More massive stars have winds that can become optically thick, where the continuum -- or part of it -- is produced in the supersonic part of the outflow.

The highest steady mass-loss rates by stellar wind are found in the late stages of the evolution of massive stars, in what is known as the Wolf--Rayet phenomenon.  Wolf--Rayet (WR) stars are very luminous stars close to the Eddington limit and characterized by highly supersonic dense winds which can shroud the hydrostatic layers from direct observation \citep{1988Langer,1989Langer,2008Crowther}. 
The mass loss is thought to be driven by radiation pressure, i.e. momentum transfer via absorption and scattering of photons in the partially optically thick wind. High momentum transfer efficiency is necessary to explain the high mass-loss rates observed in WR stars, implying that multiple scattering events and enhanced opacities due to line-blanketing are necessary to accelerate these stellar winds up to the observed terminal wind velocities of the order of thousands of km\,s$^{-1}$ \citep{1987Abbott,1999LamersCassinelli,2013Owocki,2014Bestenlehner}.

Empirical determinations of the hydrostatic radii of Galactic WR stars in recent decades has revealed a disagreement between the spectral analysis based on wind models \citep{2006Hamann,2012Sander} and stellar structure calculations of mostly helium burning stars \citep{1989Langer,1996Heger,2006Petrovic,2012Grafener,2012Georgy,2016GrassitelliWR}. 
The difficulties in reconciling this disagreement, where there is a difference  of up to a factor 10  in the derived hydrostatic radii (known as the `WR radius problem'), has been ascribed to the location of the wavelength-dependent photosphere \citep{1987Hillier,2015Hillier,1988Langer,1992Kato,1996Heger} which forms in the supersonic part of the wind. Consequently, the detailed dynamics and density structure in the expanding envelope and supersonic wind are uncertain, which relates also to the complex opacities and their interplay with the accelerating outflow \citep{1996Schaerer,1996Heger} and to the presence of density inhomogeneities in the wind \citep{1988Moffat,2009StLouis,2011CheneCIR,2014Michaux}. 

There appears to be a lack of radiative force to drive the high mass-loss rates observed in WR stars \citep[e.g.][]{2005Grafener,2015Sander}. This is true even considering the line opacity contribution of the  absorption lines (especially
UV) exposed to unattenuated radiative flux by the progressive Doppler shift of the stellar wind, which allows for the absorption of the unattenuated continuum, reducing thus the effect of line self-shadowing \citep{1970Lucy,1975CAK,1985Abbott,1993Lucy,1997Owocki}. Consequently, an approximate relation is commonly adopted to describe the dynamics of the winds (i.e. the velocity and density profiles), namely a beta-velocity law \citep{1975CAK,1989Langer,2006Hamann,2015Hillier,2017Sander}. This approximation contributes to the uncertainty in the determination of the stellar radius and the detailed temperature stratification within the \nlte\ supersonic outflow. 

Hydrostatic models of helium stars above $\sim 10 \Msun$ with Galactic metallicity show the formation of a low-density inflated envelope having a density inversion below their plane parallel  grey atmosphere \citep{2006Petrovic,2012Grafener,2016GrassitelliWR,2016Eldridge}. Inflation and density inversion arise as these models approach a local Eddington factor of unity in the proximity of the iron opacity bump at a temperature of $\approx 200$\,kK. This hydrostatic limit leads to the flattening of the temperature and density gradients in the outer layers, up to building a positive gas pressure gradient to counterbalance the strong radiative force \citep{1973Joss,1997Langer,2015Owocki,2015Sanyal}. 

 \citet{1996Heger}, \citet{1996Schaerer}, \citet{2006Petrovic}, and more recently \citet{2016Ro} and \citet{2018Nakauchi} have investigated the envelope configuration of massive stars while considering the effects of mass loss by stellar winds on the structure of the outer layers. The need for a strong radiative acceleration to launch these dense winds points towards  the role of the iron opacity bump as a source of momentum for the flow already from the subsonic quasi-hydrostatic part of the star. \citet{2006Petrovic} showed that sufficiently high mass-loss rates can lead to configurations in which the inflation of the envelope is inhibited, and where the flow is accelerated by the radiative forces in the proximity of the iron opacity bump.

In this paper we investigate the interplay between an optically thick wind and the subsonic structure of massive helium stars, focusing especially on the conditions at the sonic point. In Sect.\,2 we explore some theoretical considerations about optically thick winds and the importance of the sonic point. In Sect.\,3 we apply our considerations to computing hydrodynamic stellar structure models, in Sect.\,4 we show selected results and applications, in Sect.\,5 we develop our results to constrain predicted mass-loss rates by introducing new useful tools, in Sect.\,6 these results are discussed, and in Sect.\,7 we draw our conclusions.

\section{Optically thick winds}\label{Sect.optthickwinds}

The requirement of a smooth, radiation-pressure-driven,  steady-state, transonic flow sets specific conditions that the physical quantities and their gradients must satisfy at the sonic point for optically thick winds. 
Based on the reasonable assumption of a radiation-driven wind \citep[analogous to the O stars,][]{2015Vink}, \citet{2002Nugis} estimated that the sonic point of hot WR stars is located at high optical depth (i.e. $\tau \approx$3--30) and that it has to occur within the temperature ranges where the Rosseland opacity $\kappa$ increases as a function of radius, i.e. where the flow accelerates as a result of the increase in the local opacity ($\mathrm{d}\kappa/\mathrm{d}r>0$),  because energy and momentum input are necessary at the sonic point to accelerate the outflow. 

From the momentum equation, the velocity gradient in Eulerian coordinates describing the hydrodynamics of a steady-state stellar wind can be written as \citep{1995Bjorkman,1999LamersCassinelli,2002Nugis,2014Hubeny,2017Sander}
\begin{equation}\label{criticalpoint}
\frac{1}{\varv}\frac{\mathrm{d}\varv}{\mathrm{d}r} = -\left(g - g_\mathrm{ rad} - 2 \frac{c_\mathrm{ s}^2}{r} + \frac{\mathrm{d}c_\mathrm{ s}^2}{\mathrm{d}r}\right)/(\varv^2-c_\mathrm{ s}^2) \quad ,
\end{equation} 
where $g$ and $g_{\rm rad}$ are the gravitational and radiative accelerations, $c_\mathrm{ s}$ the local isothermal sound speed, $\varv$ the flow velocity, and $r$ the radial coordinate. Equation \ref{criticalpoint}, known as the `Bondi equation' \citep{1952Bondi}, implies that when $\varv=c_\mathrm{ s}$,  $\mathrm{d}\varv/\mathrm{d}r$ diverges unless the numerator is null as well. The requirement of a null numerator at the sonic point implies that
\begin{equation}\label{numsonic}
g_\mathrm{ rad} =  g - 2 \frac{c_\mathrm{ s}^2}{r} + \frac{\mathrm{d}c_\mathrm{ s}^2}{\mathrm{d}r}     \quad ,               
\end{equation} 
where $g=Gm/r^2$ with $m$ being the mass coordinate. In this equation the radiative acceleration can be expressed as
\begin{equation}\label{grad}
g_\mathrm{ rad} = \frac{\kappa L}{4 \pi r^2 c}        \quad ,
\end{equation}
with $\kappa$ the Rosseland mean opacity from Thompson scattering and from bound-bound and bound-free absorption, which is assumed in this context to be consistent with the flux-weighted mean opacity, and with the constants $G$ and $c$ holding their usual meaning. Assuming that the opacities are not affected by the velocity and the velocity gradient, Eq.\,\ref{criticalpoint} shows the presence of a critical point for the momentum equation at the sonic point (see Appendix \ref{appendixsoniccritical} for further discussion).    

Equation \ref{numsonic} can be written in terms of the Eddington factor $\Gamma = g_\mathrm{ rad}/g$, and becomes
\begin{equation}\label{gammasonic}
\Gamma = 1- \frac{2 \frac{c_\mathrm{ s}^2}{r} - \frac{\mathrm{d}c_\mathrm{ s}^2}{\mathrm{d}r} }{g}
,\end{equation}
which shows that the condition $\frac{\mathrm{d}c_\mathrm{ s}^2}{\mathrm{d}r} - 2 \frac{c_\mathrm{ s}^2}{r} \ll g$ leads to $\Gamma \approx 1$ (or equivalently to a local effective gravity $g_\mathrm{ eff} \approx 0$) at the sonic point. 
This has a crucial implication: for $\frac{\mathrm{d}c_\mathrm{ s}^2}{\mathrm{d}r} - 2 \frac{c_\mathrm{ s}^2}{r} \ll g$, a radiation-driven, smooth, steady-state transonic outflow from a star needs to have its sonic point located at $\Gamma\approx 1$. A transonic radiation-pressure-driven stellar wind has to self-adjust its structure to meet this requirement.

\subsection*{Temperature stratification at the sonic point  }

Formal studies of the dynamics of stellar winds and their stability in spherical symmetry were first conducted by \citet{1958Parker,1966Parker}. However, from a mathematical and conceptual point of view, solving the physical problem of the dynamics of a gas outflow is analogous to  studying  spherically symmetric mass accretion  \citep{1952Bondi,1975Tamazawa,1980Moncrief,1981Thorne,1995Bjorkman,1998Visser}. The topology of the steady-state solutions for the stellar wind equation shows two characteristic solutions: one that starts subsonic and reaches supersonic finite velocity (known as  the wind solution)  and one that starts supersonic and becomes subsonic for increasing distance (the accretion solution). 
The problem of stellar wind and accretion are conceptually analogous and, {\it mutatis mutandis}, the results obtained in one context can be applied to the other. 

Accretion onto a black hole can also be divided into optically thin and optically thick accretion depending on the location of the sonic point radius with respect to the photosphere \citep{1975Tamazawa,1981Thorne,1982Meier,1984Flammang,1991Nobili}.  In the optically thin case the hydrodynamics of the flow and the radiative transfer are effectively decoupled \citep{1978Mihalas,1981ThorneB,1982Flammang}. The non-local coupling of the radiation field to the structure of the atmosphere as a whole implies only a weak dependence of the energy transfer problem on the local thermodynamical conditions \citep{1978Mihalas}. However, as the photon mean free path $\lambda$, defined via
\begin{equation}
\lambda = (\rho \kappa)^{-1} \quad 
\end{equation}   
becomes smaller than the length scale on which the macroscopic quantities change, the system enters into the photon diffusion regime \citep{1978Mihalas,1981Thorne,1990Kippenhahn}. The radiative transport equation in the asymptotic limiting case of large optical depth and small mean free paths yields the diffusive approximation, which can be expressed as \citep{1978Mihalas,1990Kippenhahn} 
\begin{equation}\label{diffapprox}
\frac{\mathrm{d}P_\mathrm{ rad}}{\mathrm{d}\tau}=\frac{F}{c} \quad ,
\end{equation}
with $P_\mathrm{  rad}$ the radiation pressure, $F$ the flux, and $\mathrm{d}\tau = - \kappa \rho \, \mathrm{d}r$.

 Similarly, the stellar wind problem can be considered in the optically thin and optically thick regime. In Appendix \ref{SH} we investigate in more detail the behaviour of the radiation field in the proximity of the sonic point at the conditions typically met in stellar models of WNE stars. Backed by these considerations, we assume the conditions at the sonic point of WNE stars to closely match local thermodynamical equilibrium (LTE). 

The LTE assumption implies that the temperature gradient can be defined by {local} quantities only, with the energy flux set by the need to transport the stellar luminosity outwards. 
At the high optical depth assumed for the sonic point, in fact, the temperature gradient becomes independent of the detailed global structure of the atmosphere, losing its explicit dependence on the optical depth characteristic of the optically thin situation  (see Appendix B).

For a given mass-loss rate, the sonic point conditions of an optically thick wind (density and temperature) are uniquely defined,
 and thus the subsonic structure becomes independent from the detailed conditions above it, which only define the velocity profile and the terminal wind velocity of the outflow. Setting the mass-loss rate, in fact, sets the temperature at the sonic point. This can be seen combining the condition 
\begin{equation}\label{sonicpointiso}
\varv = \sqrt{\frac{k_\mathrm{ B} T}{\mu m_\mathrm{ H}}} = c_\mathrm{ s}
,\end{equation}
where $\mu$ is the mean molecular weight, $m_\mathrm{ H}$ the mass of a proton, and $k_\mathrm{ B}$ the Boltzman constant, with the definition of steady-state mass-loss rate 
\begin{equation}\label{steadymdot}
\dot{M} = 4 \pi r^2 \rho \varv
,\end{equation}
which leads to
\begin{equation}\label{tsonic}
 T_\mathrm{ S}= \frac{\mu m_\mathrm{ H}}{k_\mathrm{ B}}\left(\frac{\dot{M}}{4 \pi r^2 \rho}\right)^2 \quad .
\end{equation} 
Therefore, having the mass-loss rate imposing the blanketed sonic point temperature while coupled to the subsonic stellar structure, hydrodynamic models can be computed independently from wind calculations without the need of a global computation  (as we show in Sect.\,\ref{sectrisults}).

\subsection*{Envelope inflation}

Both hydrostatic and hydrodynamic stellar models with high luminosity-to-mass ratios close to the Eddington limit show diluted and extended envelopes, often characterized by very inefficient convective energy transport and the presence of a density inversion \citep{1997Langer,1999Ishii,2012Grafener,2015Kohler,2015Sanyal,2017Sanyal,2015GrassitelliA,2015Owocki}. These stellar models are said to be {inflated},  having a sub-photospheric envelope configuration with large pressure and density scale heights and the local Eddington factor close to unity. This happens in the temperature range where an opacity bump arises, most notably the iron opacity bump.  

Assuming that the temperature stratification is given only by radiative transport in the diffusive approximation, thus neglecting the contribution from convective energy transport \citep[appropriate for the envelopes of massive helium star models, ][]{2016GrassitelliWR}, for chemically homogeneous layers with constant luminosity, using Eq.\,\ref{diffapprox} the temperature profile can be written as
\begin{equation}\label{dTdr}
\frac{\mathrm{d}T}{\mathrm{d}r}= -\frac{\rho T}{4 P_\mathrm{  rad}}\frac{\kappa L}{4 \pi r^2 c} \quad ,
\end{equation}   
from which
\begin{equation}\label{dcsdr}
\frac{\mathrm{d}c_\mathrm{ s}^2}{\mathrm{d}r}= c_\mathrm{ s}^2\frac{\mathrm{d}\,\ln(T)}{\mathrm{d}r} =-\frac{\beta}{4(1-\beta)}\frac{\kappa L}{4 \pi r^2 c}\quad ,
\end{equation}   
with $\beta$ the ratio of gas to total pressure.
Combining then Eq.\,\ref{dTdr} with Eq.\,\ref{criticalpoint} and with the differential form of the steady-state continuity equation
\begin{equation}
\mathrm{d}\,{\ln}(\varv) + \mathrm{d}\,{\ln} (\rho) + 2\mathrm{d}\,{\ln} (r)=0\quad,
\end{equation}
the slope of the density profile follows
\begin{equation}\label{densder}
\frac{\mathrm{d}\,\ln(\rho)}{\mathrm{d}r}=\left(g-\frac{\kappa L}{4\pi r^2 c}\left(1+\frac{P_{\rm  gas}}{4P_\mathrm{  rad}}\right)+\frac{2\varv^2}{r}\right)/(\varv^2-c_\mathrm{ s}^2) \quad .
\end{equation} 
Here we can  distinguish two regimes: the subsonic and the supersonic. In the subsonic, sub-Eddington region, an increase in opacity or generally any outward-directed force (e.g. the centrifugal force) tends to reduce the slope of the density profile, thus increasing the density scale height and giving rise to the {envelope inflation} encountered  in the helium star models of e.g. \citet{2006Petrovic}, \citet{2012Grafener}, and \citet{2016GrassitelliWR}. Consequently, although counter-intuitive, from Eq.\,\ref{densder} an increase in the radiative force in the subsonic region implies a less steep increase in flow velocity, while making the density profile more and more flat; the layers aim to preserve hydrostatic equilibrium counterbalancing the local gravitational force, and the radiative acceleration effectively acts as a reduction in the local gravity. As the gas pressure gradient needed to counterbalance this reduced effective gravity is smaller, the increased opacity can lead, together with the acceleration of the flow, to an almost flat density profile when the local $\Gamma$ approaches unity \citep{2017Sanyal}. 
   
However, reaching and exceeding the local Eddington limit does not necessarily guarantee that the outflow reaches supersonic velocities. 
In case of a (still) subsonic steady-state flow, if the numerator on the right-hand side of Eq.\,\ref{densder} becomes negative, a density inversion has to arise associated with a decrease in velocity,  whereas a further increase in radiative force leads to a steep decrease in the velocity and an increase in the gas to total pressure. This hydrodynamic envelope inflation solution is similar to the `breeze' solution for winds which do not asymptotically reach supersonic velocities \citep{1999LamersCassinelli,2014Hubeny}.  

A clear definition, and therefore a criterion, of envelope inflation was  not available prior to publication of  this work \citep{2017Sanyal}. In Appendix \ref{app.inflation} we discuss more extensively the appearance of inflated envelopes. We do so by performing hydrostatic stellar structure calculations of massive helium star models using different opacities and thus comparing how the outer layers react to radiative forces of different intensity.
Non-inflated hydrostatic massive helium star models show density and gas pressure scale height which rapidly decrease as they approach the surface. This is due to the gas pressure gradient being the leading force counterbalancing gravity and preserving hydrostatic equilibrium. However, in the presence of high Eddington factors in stars with high $L/M$ ratios, an inflection point in density and gas pressure can appear, and thus density and gas pressure scale heights  start to increase. 

We show that an inflection point can appear as a consequence of a rapid increase in opacity (or equivalently in the Eddington factor) when the radiation pressure gradient becomes the primary agent counterbalancing gravity.
Analytically, an inflection point in the gas pressure profile is found when (see Eq.\,\ref{critinf}) 
\begin{equation}
\frac{\mathrm{d}\,\ln(\kappa)}{\mathrm{d}\,\ln(T)} \lesssim 1 -\frac{1}{\Gamma} \quad, 
\end{equation}
showing both the need for an Eddington factor close to unity and for a steep increase in the opacity with temperature.
In view of this discussion and of Appendix \ref{app.inflation}, we can thus understand envelope inflation as {the appearance of large gas pressure scale heights in response to a steep increase in the radiative opacity in radiation-pressure-dominated layers close to the Eddington limit}. 
The low gas pressure gradient prevents the density from decreasing steeply for extended regions of space, and implies that the surface is found at significantly larger radii compared to non-inflated stellar models. 

\section{Model assumptions}\label{model}

Computing the stellar structure is a task that requires solving a hyperbolic set of differential equations simultaneously with well-defined boundary conditions \citep{1990Kippenhahn}.
We adopt the Bonn evolutionary code (BEC), a Lagrangian {\it hydrodynamic} one-dimensional stellar evolution code \citep{1988Langer,1994Langer,1996Heger,1998Heger,2000Heger,2006Petrovic,2006Yoon,2014Kozyreva}.
It solves the set of coupled non-linear partial differential stellar structure equations in the form \citep{1990Kippenhahn}

\begin{equation}\label{consmass}
\left(\frac{\partial m}{\partial r}\right)_t = 4 \pi r^2 \rho
,\end{equation}
\begin{equation}\label{defvel}
 \left(\frac{\partial r}{\partial t}\right)_m = \varv\\
,\end{equation}
\begin{equation}\label{momentumeq}
 \left(\frac{a}{4 \pi r^2}\right)_t = \frac{G m}{4 \pi r^4} + \frac{\partial P}{\partial m}\\
,\end{equation}
\begin{equation}\label{energytransp}
 \left(\frac{\partial T}{\partial m}\right)_t = - \frac{G m }{4 \pi r^4 }\frac{T}{P}\nabla \left( 1+ \frac{r^2}{G m}\frac{ \partial \varv}{\partial t}  \right)_m
,\end{equation}
\begin{equation}\label{energycons}
 \left(\frac{\partial L}{\partial m}\right)_t=\epsilon_n- \epsilon_g - \epsilon_\nu
,\end{equation}
where $m$ and $t$, the mass coordinate and  time, are the two independent variables;  $\rho$ is the density; $\varv$ is the velocity; $a$ is the acceleration; $r$ is the radial coordinate; $T$ is the temperature; $L$ is the luminosity; $\nabla:=\frac{\mathrm{d}\,\log(T)}{\mathrm{d}\,\log(P)}$ is the temperature gradient; $P$ is the total pressure given by the sum of gas and radiation pressure; $G$ is the gravitational constant; and $\epsilon$ is the energy production/loss per unit mass and unit time related to nuclear processes (subscript n), to gravitational contraction/expansion (subscript g), and neutrino loss (subscript $\nu$).   
These equations express: the conservation of mass (Eq.\,\ref{consmass}), the definition of velocity (Eq.\,\ref{defvel}), the conservation of momentum  (Eq.\,\ref{momentumeq}), the energy transport  (Eq.\,\ref{energytransp}), and the energy conservation  (Eq.\,\ref{energycons}). These equations together with the network of nuclear reaction rates, the set of equations of the mixing length theory \citep{1958Vitense}, and the OPAL opacity tables \citep{1996Iglesias}, define the structure and evolution of a stellar model. In addition, mass loss by stellar wind can also be applied.   
  
The set of non-linear coupled partial differential equations above has five dependent variables (namely $\rho$, $L$, $\varv$, $T$, and $r$) and requires a number of boundary conditions equal to the degrees of freedom of the system, i.e. five. The central boundary conditions are trivially set by the requirement of zero $r$, $L$, and $\varv$ in the centre of the stellar models, while the outer boundary conditions are usually determined by the assumption of plane parallel grey atmosphere \citep{1989Langer,1998Heger}. 

The boundary conditions set constraints on the family of possible solutions (if any) of the set of partial differential equations describing a star. We modify BEC adopting as surface boundary conditions
\begin{equation}\label{boundmdot}
\dot{M} = 4 \pi r^2 \rho \varv
\end{equation}
and
\begin{equation}\label{sonicpoint}
\varv = \sqrt{\frac{k_\mathrm{ B} T}{\mu m_\mathrm{ H}}} = c_\mathrm{ s}
,\end{equation}
i.e. we impose the boundary of the stellar model at the sonic point while preserving continuity. 
The mass-loss (or mass-accretion) rate is a free parameter in our calculations,  i.e. it can be chosen either manually or according to mass loss by stellar wind prescriptions \citep[such as those in e.g.][]{2000Nugis,2001Vink}. Mass loss is treated adopting a pseudo-Lagrangian scheme in the outer part of the stellar model, i.e. the independent variable becomes $q=m/M_\mathrm{ tot}$ (the relative mass coordinate) and consequently the Lagrangian operator time derivative for a steady state becomes $\frac{\mathrm{d}}{\mathrm{d}t}= - \frac{q \dot{M}}{M_\mathrm{ tot}}\frac{\partial}{\partial q} $ \citep{1977Neo,1998Heger}.

We adopt BEC to compute the subsonic structure of massive helium stars. Models are computed with a chemical composition as in \citet{2000Heger}, with metallicity Z=0.02 and helium fraction 0.98, and stationarity is achieved with large time steps (i.e. greater than $10^3$ seconds) while inhibiting chemical evolution. 
The velocity profile is computed self-consistently with the radiative acceleration derived via the velocity-independent Rosseland opacity and the imposed mass-loss rate. Convection is included in the calculations, but as shown by \citet{2016GrassitelliWR}, the convective flux in the outer layers is several orders of magnitude smaller than the radiative flux. 
We compute models for chemically homogeneous massive helium zero age main sequence stars  of mass 10, 15, and 20 \Msun at solar metallicity and under various applied mass-loss rates by stellar wind. Rotation and magnetic field are not included in the calculations; however, they are not expected to affect the general theoretical considerations in Sect.\,2.
We also neglect the structural effects from the turbulent pressure on the stellar models.
 For comparison purposes, we also compute models with classical plane parallel grey atmosphere boundary conditions \citep[cf. ][]{1998Heger,2006Petrovic}.

\section{Results}\label{sectrisults}

\begin{table*}
{\caption{
 Sonic point properties of our set of chemically homogeneous helium star models with a mass of 15\Msun (see Sect.\,\ref{sectrisults}).
 The listed quantities are, starting from the left: mass-loss rate applied, sonic point radius, luminosity, Rosseland opacity, density, temperature, photon mean free path, flow velocity, Eddington factor, optical depth parameter, pressure scale height, effective temperature computed via the Stefan--Boltzmann law, optical depth. All these quantities are given at the sonic point.     
}\label{tab}}
\centering

\begin{tabular}{ c c c c c c c c c c c c c }
\hline\hline
  $\log\, (\dot{M})$ & $R_\mathrm{ S}$  & $\log\, (L) $ & $\kappa$ &  $\log\, (\rho_\mathrm{ S})$  & $\log\, (T_\mathrm{ S})$ & $\log\, (\lambda) $  & $\varv$ & $\Gamma$ & $t$ & $\log\,(H_P)$ & $\log\, (T_\mathrm{ eff})$ & $\tau_\mathrm{ S}$  \\
    \hline     
    $M_\odot/$yr & $R_\odot$ & $L_{\odot}$ & cm$^2$/g & g/cm$^3$ & K & cm & km/s &$ $ & $ $ & cm & K & $ $ \\ 
\hline 
% -4.3 & 1.152 & 5.460 & 0.6953 & -7.967 & 5.324 & 8.125 & 36.23 & 1.016 & 18.31 & 9.185 & 5.096 & 33 \\
% -4.52 & 1.159 & 5.463 & 0.6834 & -8.185 & 5.304 & 8.350 & 35.41 & 1.010 & 15.97 & 9.324 & 5.095 & 19 \\
% -4.7 & 1.167 & 5.465 & 0.6856 & -8.357 & 5.285 & 8.521 & 34.63 & 1.011 & 10.49 & 9.423 & 5.094 & 13 \\
% -5.0 & 1.184 & 5.467 & 0.6739 & -8.655 & 5.254 & 8.827 & 33.43 & 1.006 & 8.564 & 9.609 & 5.092 & 7 \\
% -5.1 & 1.194 & 5.467 & 0.6773 & -8.752 & 5.239 & 8.921 & 32.86 & 1.005 & 7.340 & 9.652 & 5.090 & 5 \\
% -5.3 & 1.219 & 5.468 & 0.6682 & -8.958 & 5.205 & 9.133 & 31.61 & 1.001 & 10.79 & 9.742 & 5.086 & 3 \\
% -5.4 & 1.260 & 5.470 & 0.7048 & -8.833 & 4.705 & 8.985 & 17.76 & 1.018 & 0.203 & 7.748 & 5.079 & 2 \\
 
 -4.3 & 1.152 & 5.464 & 0.658 & -7.971 & 5.331 & 8.125 & 36.55 & 0.9901 & 8.097 & 9.218 & 5.099 & 33 \\
 -4.5 & 1.158 & 5.466 & 0.659 & -8.165 & 5.312 & 8.350 & 35.73 & 0.9918 & 6.689 & 9.335 & 5.097 & 20 \\
 -4.7 & 1.165 & 5.467 & 0.660 & -8.359 & 5.291 & 8.521 & 34.91 & 0.9931 & 5.754 & 9.451 & 5.096 & 13 \\
 -5.0 & 1.182 & 5.469 & 0.661 & -8.656 & 5.259 & 8.827 & 33.62 & 0.9946 & 4.708 & 9.628 & 5.093 & 8 \\
 -5.1 & 1.190 & 5.468 & 0.661 & -8.753 & 5.246 & 8.921 & 33.14 & 0.9952 & 4.787 & 9.680 & 5.091 & 6 \\
 -5.3 & 1.216 & 5.469 & 0.661 & -8.959 & 5.212 & 9.133 & 31.87 & 0.9964 & 5.790 & 9.768 & 5.087 & 3 \\
 -5.4 & 1.260 & 5.470 & 0.626 & -8.843 & 4.726 & 8.985 & 18.20 & 0.8978 & 0.138 & 7.827 & 5.079 & 2 \\ 
 \hline

\end{tabular}

\vspace{0.4cm}

\centering

Same as above, but for our set of models with a mass of 10\Msun 
\vspace{0.1cm}
 
\begin{tabular}{ c c c c c c c c c c c c c }
\hline\hline
  $\log\, (\dot{M})$ & $R_\mathrm{ S}$  & $\log\, (L) $ & $\kappa$ &  $\log\, (\rho_\mathrm{ S})$  & $\log\, (T_\mathrm{ S})$ & $\log\, (\lambda) $  & $\varv$ & $\Gamma$ & $t$ & $\log\,(H_P)$ & $\log\, (T_\mathrm{ eff})$ & $\tau_\mathrm{ S}$  \\
    \hline     
    $M_\odot/$yr & $R_\odot$ & $L_{\odot}$ & cm$^2$/g & g/cm$^3$ & K & cm & km/s &$ $ & $ $ & cm & K & $ $ \\ 
\hline 
% -4.3 & 0.912 & 5.129 & 0.9804 & -7.750 & 5.295 & 7.759 & 35.06 & 1.009 & 37.01 & 8.840 & 5.064 & 49 \\
% -4.52 & 0.916 & 5.131 & 0.9712 & -7.961 & 5.264 & 7.974 & 33.84 & 1.006 & 28.99 & 8.925 & 5.064 & 26 \\
% -4.7 & 0.922 & 5.133 & 0.9602 & -8.124 & 5.228 & 8.141 & 32.46 & 0.999 & 34.12 & 8.946 & 5.063 & 16 \\
% -5.0 & 0.930 & 5.136 & 0.9785 & -8.213 & 4.788 & 8.222 & 19.55 & 0.966 & 0.883 & 7.485 & 5.062 & 8 \\
% -5.1 & 0.930 & 5.137 & 0.9919 & -8.299 & 4.766 & 8.302 & 19.05 & 0.965 & 0.754 & 7.474 & 5.062 & 6 \\
% -5.3 & 0.931 & 5.137 & 1.0200 & -8.481 & 4.722 & 8.472 & 18.10 & 1.000 & 0.366 & 7.461 & 5.062 & 3 \\

 -4.3 & 0.910 & 5.115 & 0.986 & -7.748 & 5.295 & 7.748 & 35.03 & 0.9828 & 9.766 & 8.832 & 5.061 & 49 \\
 -4.5 & 0.914 & 5.120 & 0.979 & -7.937 & 5.267 & 7.937 & 33.93 & 0.9861 & 8.418 & 8.908 & 5.061 & 28 \\
 -4.7 & 0.920 & 5.123 & 0.971 & -8.118 & 5.221 & 8.118 & 32.17 & 0.9883 & 13.29 & 8.909 & 5.061 & 16 \\
 -5.0 & 0.930 & 5.136 & 0.842 & -8.225 & 4.814 & 8.225 & 20.13 & 0.8173 & 0.625 & 7.565 & 5.061 & 8 \\
 -5.1 & 0.930 & 5.136 & 0.846 & -8.311 & 4.791 & 8.311 & 19.62 & 0.8082 & 0.523 & 7.556 & 5.061 & 6 \\
 -5.3 & 0.931 & 5.136 & 0.855 & -8.493 & 4.747 & 8.493 & 18.64 & 0.7866 & 0.377 & 7.544 & 5.061 & 3 \\
\hline
\end{tabular}

\vspace{0.4cm}
\centering
 Same as above, but for our set of models with a mass of 20\Msun 
\vspace{0.1cm}

\begin{tabular}{ c  c  c  c  c  c  c  c  c  c  c  c  c }
%\vspace{0.2cm}
\hline\hline              
  $\log\, (\dot{M})$ & $R_\mathrm{ S}$  & $\log\, (L) $ & $\kappa$ &  $\log\, (\rho_\mathrm{ S})$  & $\log\, (T_\mathrm{ S})$ & $\log\, (\lambda) $  & $\varv$ & $\Gamma$ & $t$ & $\log\,(H_\mathrm{ P})$ & $\log\, (T_\mathrm{ eff})$ & $\tau_\mathrm{ S}$  \\
    \hline     
    $M_\odot/$yr & $R_\odot$ & $L_{\odot}$ & cm$^2$/g & g/cm$^3$ & K & cm & km/s &$ $ & $ $ & cm & K & $ $  \\ 
\hline
% -4.3 & 1.364 & 5.678 & 0.5581 & -8.122 & 5.341 & 8.376 & 36.95 & 1.016 & 13.03 & 9.425 & 5.114 & 27 \\
% -4.52 & 1.374 & 5.679 & 0.5552 & -8.342 & 5.322 & 8.597 & 36.17 & 1.013 & 9.419 & 9.575 & 5.113 & 16 \\
% -4.7 & 1.385 & 5.680 & 0.5528 & -8.517 & 5.308 & 8.775 & 35.56 & 1.010 & 8.020 & 9.697 & 5.111 & 11 \\
%-5.0 & 1.412 & 5.681 & 0.5501 & -8.822 & 5.281 & 9.081 & 34.48 & 1.006 & 6.444 & 9.909 & 5.107 & 6 \\
% -5.1 & 1.424 & 5.682 & 0.5494 & -8.922 & 5.272 & 9.182 & 34.13 & 1.005 & 5.858 & 9.980 & 5.105 & 2 \\
% -5.3 & 1.458 & 5.682 & 0.5476 & -9.137 & 5.252 & 9.398 & 33.37 & 1.003 & 5.200 & 10.14 & 5.100 & 1 \\

 -4.3 & 1.362 & 5.678 & 0.545 & -8.123 & 5.344 & 8.376 & 37.09 & 0.9934 & 5.158 & 9.438 & 5.114 & 27 \\
 -4.5 & 1.371 & 5.679 & 0.544 & -8.320 & 5.327 & 8.597 & 36.37 & 0.9945 & 4.213 & 9.571 & 5.113 & 18 \\
 -4.7 & 1.383 & 5.680 & 0.543 & -8.517 & 5.310 & 8.775 & 35.68 & 0.9954 & 3.635 & 9.707 & 5.111 & 11 \\
 -5.0 & 1.409 & 5.681 & 0.542 & -8.821 & 5.283 & 9.081 & 34.58 & 0.9963 & 2.931 & 9.917 & 5.108 & 6 \\
 -5.1 & 1.421 & 5.681 & 0.542 & -8.921 & 5.274 & 9.182 & 34.23 & 0.9965 & 2.731 & 9.988 & 5.106 & 2 \\
 -5.3 & 1.453 & 5.682 & 0.542 & -9.135 & 5.255 & 9.398 & 33.48 & 0.9971 & 2.487 & 10.14 & 5.101 & 1 \\
\hline
\end{tabular}

\end{table*}

\begin{figure}[b]
\resizebox{\hsize}{!}{\includegraphics{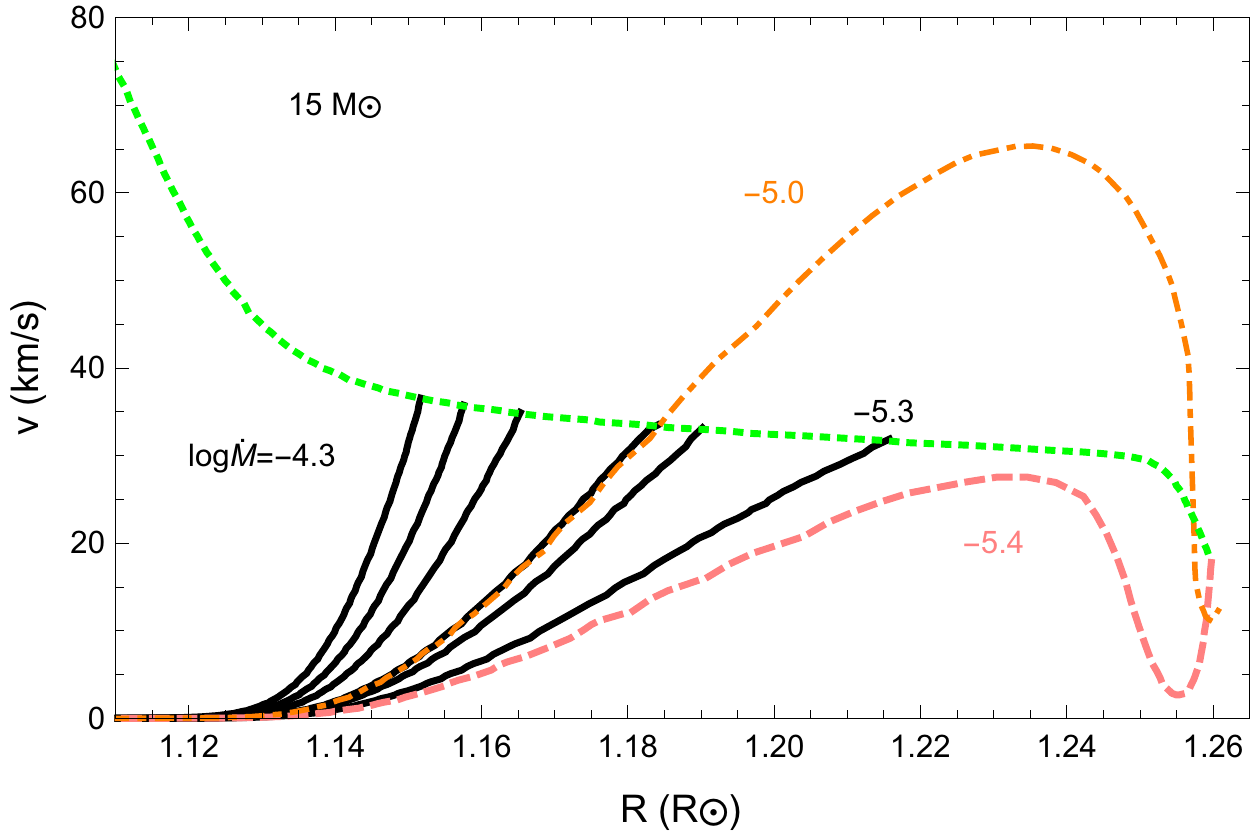}}
\caption{Velocity as a function of radius for a set of 15\Msun stellar models with different mass-loss rates and boundary conditions. The black continuous lines (from right to left) indicate models with $\Mdot$ equal to  5, 8, 10, 20, 30, 50 $\times 10^{-6} \Msun\,{\rm yr}^{-1}$ and the pink dashed line indicates the model with $\Mdot =4\times 10^{-6} \Msun\,{\rm yr}^{-1}$, all having sonic point boundary conditions. The orange dot-dashed lines indicates the model with $\Mdot =1\times 10^{-5} \Msun\,{\rm yr}^{-1}$ computed with plane parallel grey atmosphere boundary conditions. The green dashed line indicates the local isothermal sound speed from the pink dashed model.
}
\label{env}
\end{figure}

\begin{figure}
\resizebox{\hsize}{!}{\includegraphics{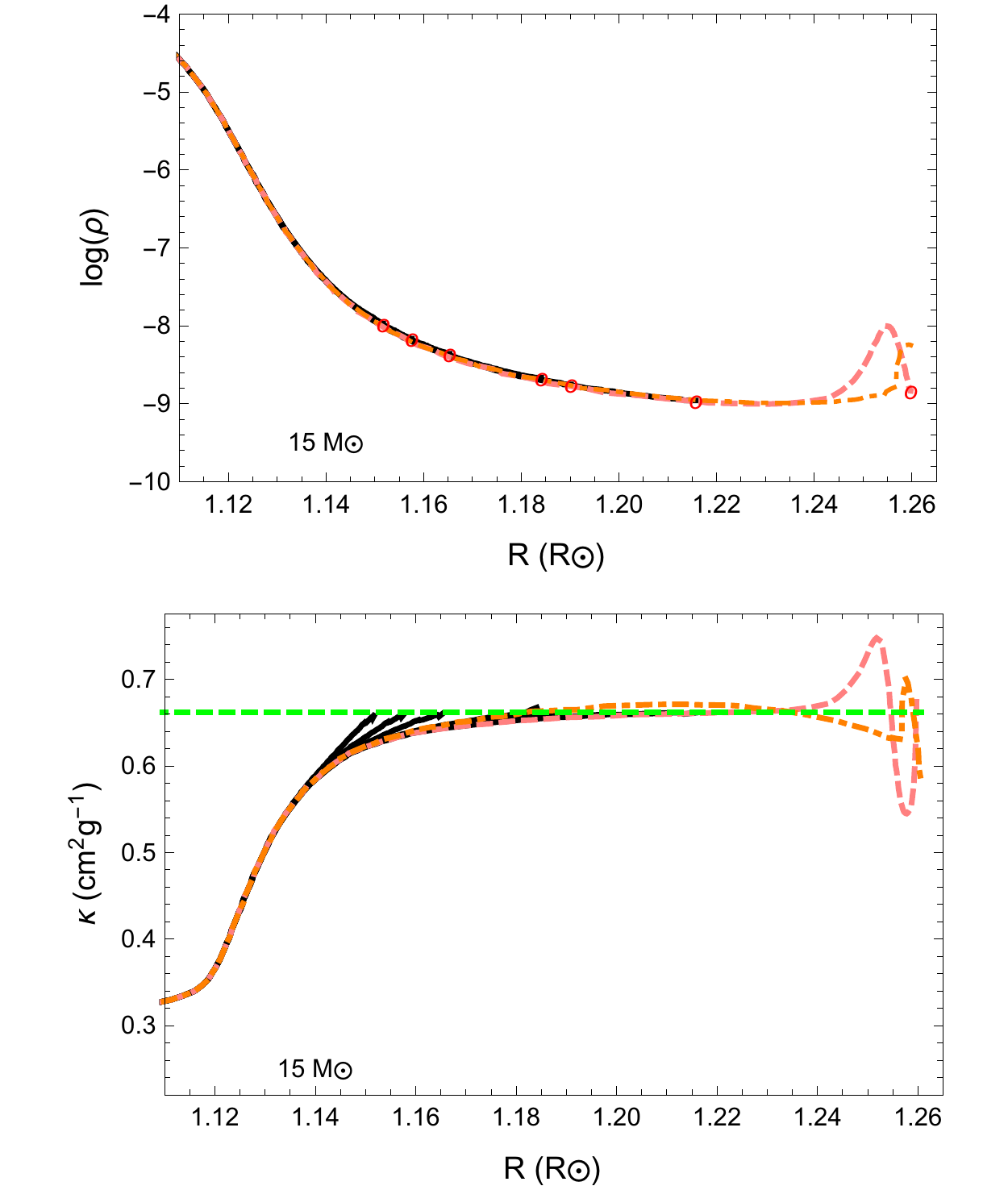}}
\caption{Density profile (top, in units of g/cm$^3$) and opacity profile (bottom, in units of cm$^2$/g) as a function of radius for the same 15\Msun stellar models with different mass-loss rates and boundary conditions as in Fig.\,\ref{env}. The colours are as in Fig.\,\ref{env}, and the red circles indicate the location of the sonic point and the green dashed line the Eddington opacity (as defined in Eq.\,\ref{kedd} in Sect.\,\ref{sectrisults}) of the pink dashed model.
}
\label{den}
\end{figure}

%We compute the subsonic structure of massive helium star models by means of hydrodynamic stellar structure calculations and investigate the detailed conditions at the sonic point of our set of stellar models. 
Figure\,\ref{env} shows the velocity profiles of the outflow in the outer subsonic part of a set of 15\Msun stellar models computed with sonic point boundary conditions and with different adopted mass-loss rates. For comparison the velocity profile of a plane parallel grey atmosphere BEC model is also shown. 

All the models presented in Fig.\,\ref{env} show a rapid increase in the radial velocity starting from  $R \approx 1.13\, R_\odot$. This increase is associated with the increase in radiative force in correspondence with the iron opacity bump (Fe-bump) in the temperature range $5.2\lesssim {\log(T)}\lesssim 5.5$. The more compact, black models in Fig.\,\ref{env} show that the velocity profile is steeper for the higher mass-loss rate applied, with a flow velocity monotonically increasing until it reaches the sound speed in the temperature range of the hot rising part of the Fe-bump. The model with the highest mass-loss rate (5$\times 10^{-5}\Msun\,{\rm yr}^{-1}$) achieves a transonic flow from the smallest radius, while the other stellar models show larger sonic point radii as the adopted mass-loss rates decrease (see Table \ref{tab} for the parameters of the models). All the monotonic solutions are thus solutions where the outflow is accelerated by the iron opacity (see also Fig.\,\ref{den}). Compared to the electron scattering opacity, this increase in opacity provides a stronger outward-directed radiative force, leading to higher local Eddington factors and flows that can reach supersonic velocities already at 200\,kK. The higher the mass-loss rate, the steeper the velocity profile,  the sooner (in terms of radius) the flow meets the sound speed, and the hotter and more compact  is the resulting model.
 %Owing to the boundary conditions, in fact, from Eq.\,\ref{sonicpoint}, the velocity profiles are such that they meet the condition $\varv=c_\mathrm{ s}$ at 
%\begin{equation}\label{tsonic}
% T_\mathrm{ S}= \frac{\mu m_\mathrm{ H}}{k_\mathrm{ B}}\left(\frac{\dot{M}}{4 \pi r^2 \rho}\right)^2 \quad ,
%\end{equation}
%with therefore increasing sonic point temperature (and isothermal sound speed) for increasing mass-loss rates. 

 However, solutions with non-monotonic velocity profiles are also present. While the group of more compact monotonic solutions has mass-loss rates of $\Mdot\geq 5\times10^{-6}\Msun\,{\rm yr}^{-1}$ (see Table \ref{tab}), the pink model with a mass-loss rate of $\Mdot = 4\times10^{-6}\Msun\,{\rm yr}^{-1}$ shows an outflow that does not become supersonic in the proximity of the iron opacity bump. In this case the flow slows down to velocities of the order of a few km/s after being accelerated by the increased opacity of the Fe-bump. It then re-accelerates and finds the sonic point in the hot rising part of the helium opacity bump (He-bump)  when $\log(T)\approx 4.6-5$. This kind of extended solution implies the formation of a density inversion following the decrease in flow velocity. This solution approaches and exceeds $\Gamma=1$ in the envelope (in correspondence with the positive density gradient), but shows an outflow that does not exceed the local sound speed before the opacity peak of the Fe-bump (${\log(T)}\approx 5.2$). For the 15\Msun model, however, even in these more extended solutions, the sonic point radius does not increase by more than 10\% compared to the more compact solutions.
 
For comparison, a stellar model with plane parallel grey atmosphere boundary conditions and a mass-loss rate of $10^{-5}\Msun\,{\rm yr}^{-1}$ is plotted in Fig.\,\ref{env}, which shows an outflow accelerating to supersonic velocities near the Fe-bump, but then decelerating to give rise to a density inversion below the peak temperature of the iron opacity bump. This model also shows that the subsonic structure of the model computed with plane parallel grey atmosphere boundary conditions and the one with sonic point boundary conditions do not differ. This is the case even if the first has layers above the sonic point while the second has no information concerning the conditions above the sonic point other than the prescribed mass-loss rate.

Figure \ref{den} shows the density inversion in the plane parallel and He-bump solutions, with a peak density one order of magnitude higher than the underlying layers. The inflection point in the density profile (i.e. $\mathrm{d}^2\rho/\mathrm{d}r^2=0$) is also visible at $R\approx 1.13 \Rsun$, indicating the location where inflation begins (see Sect.\,\ref{Sect.optthickwinds}). This radius is  defined as the core radius. For the compact, Fe-bump solutions, the larger the radius, the lower the density of the sonic point, and the larger the density scale height. Figure \ref{den} also shows the steep increase in opacity due to the recombination of iron encountered at the base of envelope. This Fe-bump increases the opacity with respect to the electron scattering opacity by almost a factor of 3, leading to the appearance of the inflection point and providing the momentum to accelerate the flows up to the sonic point from temperatures as high as a  few hundred thousand kelvins. 

%Equation \ref{kedd} can help us rewriting the radiative energy transport in the form
%\begin{equation}
%\frac{dP_\mathrm{  rad}}{dP}= \frac{\kappa}{\kappa_{EDD}} \quad ,
%\end{equation}  
%indicating that the pressure gradient in the proximity of the Eddington limit is mostly defined by the radiation pressure gradient. 
%It is also visible how instead the layers with an increase in density are super-Eddington, having opacity which exceed the Eddington opacity.    

As a consequence of the need for a $\mathrm{d}\kappa/\mathrm{d}r>0$ \citep{2002Nugis}, for the 15\Msun helium star models we find two types of configurations depending on the applied mass-loss rate: 
\begin{itemize}
\item  hot, Fe-bump stellar models having corresponding sonic temperatures which are higher than $\log(T_\mathrm{ S}) \gtrsim 5.2$ and radii similar to the core-size;
\item cooler, slightly more extended He-bump models with sonic points in the range $\log(T_\mathrm{ S})\approx 4.6-4.8$. 
\end{itemize} 
   
Similarly, in Fig.\,\ref{env10} we show how a 10\Msun helium star model readjusts to the different mass-loss rates. As in Fig.\,\ref{env}, higher mass-loss rates lead to more compact solutions, with steeper velocity profiles. Differently from the 15\Msun model, the more compact and less luminous 10\Msun helium star models find the sonic point in the helium opacity bump already starting from $ \Mdot = 1\times10^{-5}\Msun\,{\rm yr}^{-1}$ (see Table \ref{tab}). 
This is due to the lower luminosities and higher densities in the envelopes of these models.  

The opposite is true for the 20\Msun model in Fig.\,\ref{env20}, for which the solutions reach the sonic point in the hot part of the Fe-bump at $\approx 1.4 R_{\odot}$ for all the applied mass-loss rates. In this case the high luminosities (and low densities in the envelope) of the models allow a transonic flow, even for relatively low mass-loss rates. 
Figure \ref{env20} also shows how the model with a plane parallel grey atmosphere behaves in the supersonic part of the outflow (for $\rm \Mdot = 8\times10^{-6}\Msun\,{\rm yr}^{-1}$), which rapidly reaches flow velocities of more than 100 km/s, to then slow down again following the density inversion due to the decrease in opacity between the Fe- and He-bump.     

\subsection*{Sonic point conditions}

In Table \ref{tab}, the physical quantities computed at the sonic point are reported for our stellar models. The mean free paths estimated at the sonic point are small, of the order of 0.01--1\% of the sonic radius and an order of magnitude smaller than the pressure scale height at the sonic point. They tend to increase for the lower mass loss by stellar wind applied and for the He-bump solutions, mostly due to the lower sonic point densities. 
From Appendix \ref{SH} and the analysis concerning the radiation field at the sonic point, mean free paths smaller than the scale at which the thermodynamic quantities change suggest that local conditions are not far from LTE. 

For the diffusive approximation to be valid, the mean free path should also be considered together with the optical depth of the sonic point. We estimate the optical depth of the wind via the integral
\begin{equation}
\tau_\mathrm{ S}=\int_{R_\mathrm{ S}}^{\infty} \mathrm{d}r \,\, \kappa(\rho,T) \,\rho
\end{equation}   
from the sonic point radius $\rm R_\mathrm{ S}$ to infinity. In the supersonic region $\rho$ is computed via the continuity equation (Eq.\,\ref{boundmdot}) while adopting a beta-velocity law \citep[][Eq.\,11]{1989Langer} connected continuously on top of the stellar model, having the exponent unity and a typical velocity at infinity of 1600 km/s. The opacity $\kappa$ is instead computed via the OPAL opacity tables throughout the wind. To evaluate the opacity, the OPAL tables need as input not only the local density, but also the temperature, approximated in this case via the differential form of the $T-\tau$ relation from \citet{2002Nugis}  with the sonic point temperature as reference temperature. These values, which should  be considered only a rough estimate of the actual optical depth, are shown in Table \ref{tab} and  indicate that the optical depth of the sonic point is of the order of 10 in case of mass-loss rates greater than $\log(\Mdot)\gtrsim -5$. Instead, for $\log(\Mdot)\lesssim -5$  the optical depth becomes close to unity and the mean free path at the sonic point becomes quite large, indicating that the assumption of diffusive energy transport and LTE might not be valid for these mass-loss rates. 

Another key difference between the He-bump and the Fe-bump models can be seen considering the optical depth parameter \citep{1975CAK,1978Mihalas,1999LamersCassinelli}  
\begin{equation}
t = \frac{\varv_\mathrm{ th}}{\lambda}/\frac{\mathrm{d}\varv}{\mathrm{d}r}
,\end{equation}
with $\varv_\mathrm{ th}$ being the thermal velocity of the particles \citep{1999LamersCassinelli}. The optical depth parameter tends to be of the order of 10 in the hot Fe-bump solutions, while it drops to less than 1 in the He-bump case. Considering that the optical depth parameter indicates how well the velocity independent Rosseland mean opacity can reproduce the flux-weighted opacity also in the presence of a velocity gradient, low values of $t$ point to the inadequacy of the use of the OPAL opacity tables in this context. For low $t$, the radiative acceleration due to the progressive line-deshadowing associated with the velocity gradient could instead already be important for launching these winds.   

\begin{figure}
\resizebox{\hsize}{!}{\includegraphics{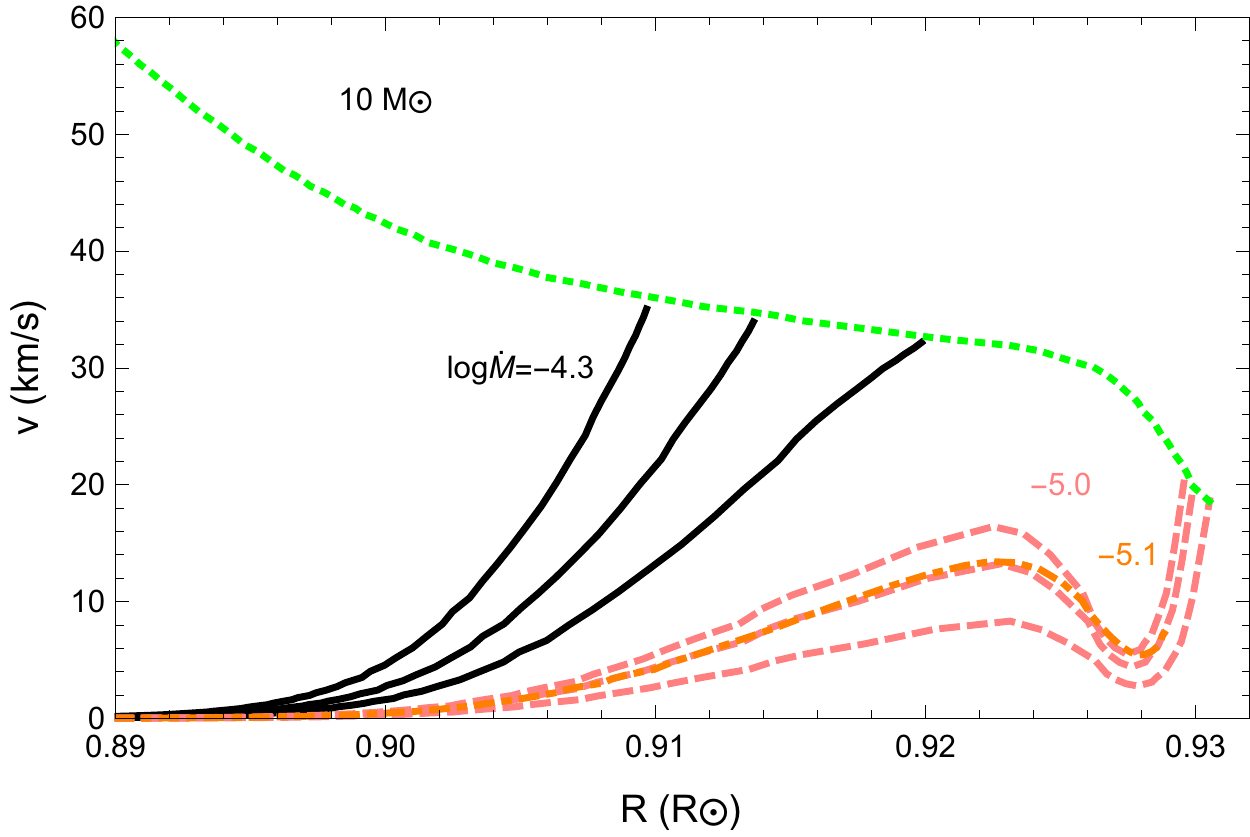}}
\caption{Velocity profile as a function of radius for a set of 10\Msun stellar models with different mass-loss rates (as in Fig.\,\ref{env}). The black continuous lines indicate models with $\Mdot \geq 2 \times 10^{-5} \Msun\,{\rm yr}^{-1}$ (2, 3, 5$\times 10^{-5}$), while the pink dashed lines indicate models with $\Mdot \leq 1 \times 10^{-5} \Msun\,{\rm yr}^{-1}$ (5, 8, 10$\times 10^{-6}$). The green dashed line and the orange dot-dashed line indicate the local isothermal sound speed of the model with $\Mdot = 5 \times 10^{-6} \Msun\,{\rm yr}^{-1}$ and the plane parallel grey atmosphere model with $\Mdot = 8 \times 10^{-6} \Msun\,{\rm yr}^{-1}$, respectively. }
\label{env10}
\end{figure}

\begin{figure}
\resizebox{\hsize}{!}{\includegraphics{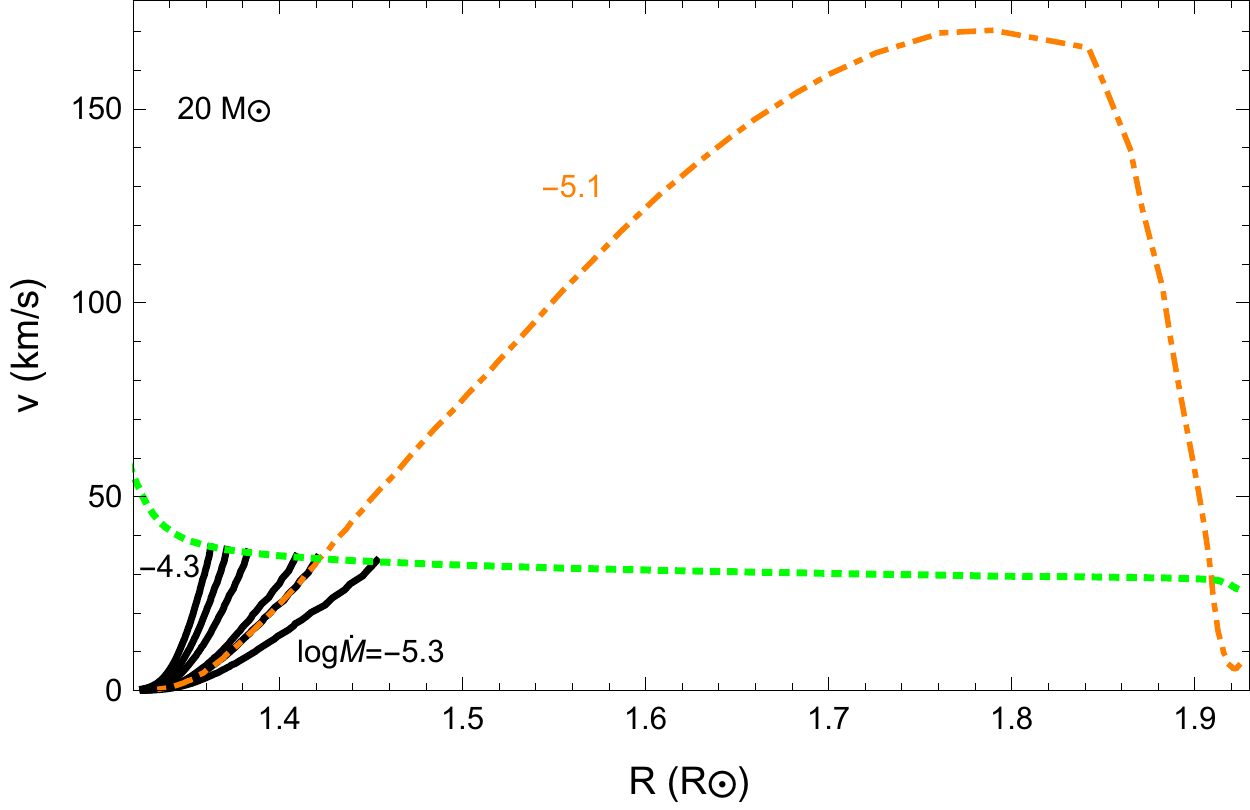}}
\caption{Velocity profile as a function of radius for a set of 20\Msun stellar models with different mass-loss rates (as in Fig.\,\ref{env}). The black continuous lines indicate models with $\Mdot \geq 5 \times 10^{-6} \Msun\,{\rm yr}^{-1}$ (5, 10, 20, 30, 50$\times 10^{-6}$), while the orange dot-dashed lines indicate models with $\Mdot = 8\times 10^{-6} \Msun\,{\rm yr}^{-1}$ computed with plane parallel grey atmosphere. The green dashed line indicates the local isothermal sound speed.
}
\label{env20}
\end{figure}

\begin{figure*}[!]
\begin{sidecaption}
\resizebox{0.7\hsize}{!}{
\includegraphics{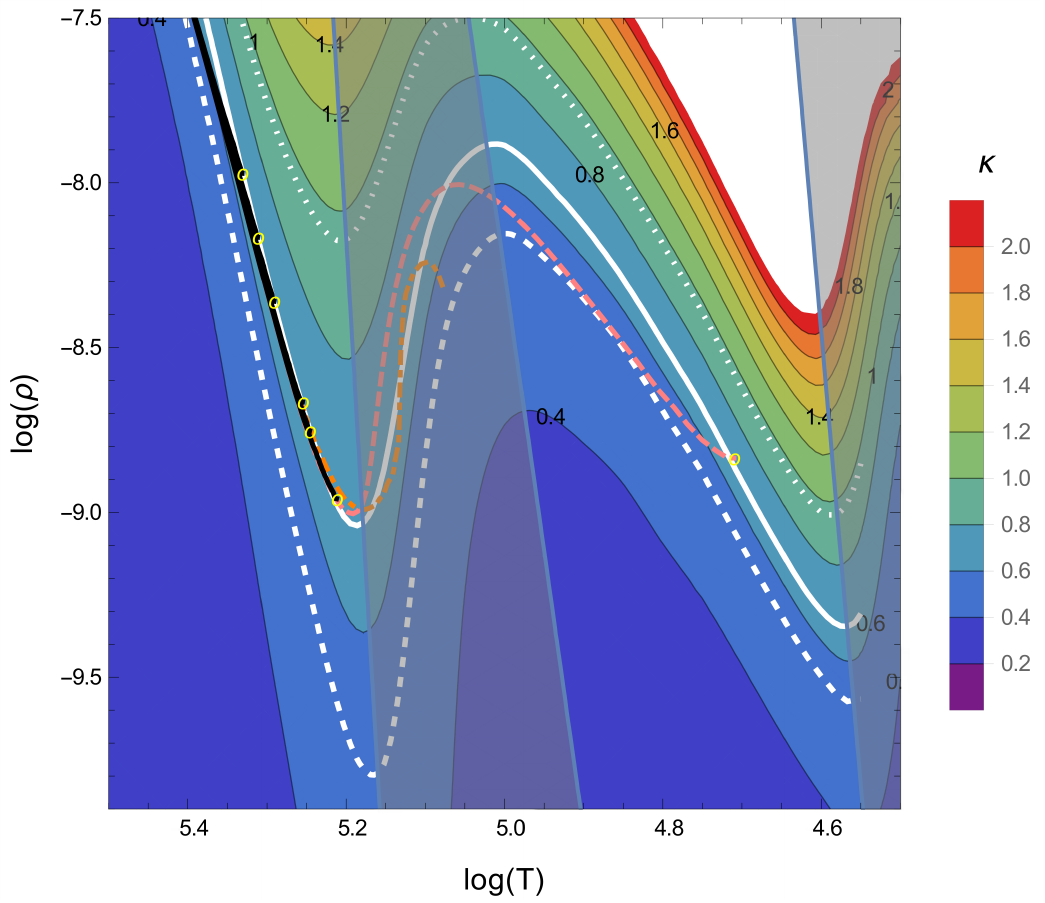}}
\caption{Contour plot showing the opacity (in cm$^2$/g) from OPAL tables (colour-coded, see scale at right) as a function of temperature (in K) and of density (in g/cm$^3$) for a chemical composition of [X,Y,Z]=[0,0.98,0.02], with X hydrogen mass fraction, Y helium mass fraction, and Z metallicity. The black lines correspond to the outer layers of the hot, more compact 15$\Msun$ stellar models with $\Mdot \geq 8\times 10^{-6} \Msun\,{\rm yr}^{-1}$ from Fig.\,\ref{env} (see also Table \ref{tab}), while the pink line indicates the cooler model with $\Mdot = 4\times 10^{-6} \Msun\,{\rm yr}^{-1}$, and the orange dot-dashed line the plane parallel grey atmosphere model with $\Mdot = \times 10^{-5} \Msun\,{\rm yr}^{-1}$. The yellow circles indicate the locations of the sonic points of this set of models, while the white lines indicate the iso-contours of the Eddington opacity for the 10\Msun (dotted), 15\Msun (continuous), and 20\Msun (dashed) stellar models. Grey areas indicate regions in the $\log(\rho)-\log(T)$ diagram where the sonic point of a radiation-driven wind cannot be located ($\mathrm{d}\kappa/\mathrm{d}r \leq 0$). 
}\label{tro}
\end{sidecaption}
\end{figure*}
 
A key aspect that emerges from Table \ref{tab} is that the sonic point coincides to a very good approximation with an Eddington factor of unity.
In general the Eddington limit can be rewritten in the form of an Eddington opacity $\kappa_\mathrm{ EDD}$,
\begin{equation}\label{kedd}
\kappa_\mathrm{ EDD}:= 4 \pi c G M / L \quad,
\end{equation}  
only proportional to the mass-to-luminosity ratio; this is the opacity necessary for a star in hydrostatic equilibrium to reach the Eddington limit. 

Figure \ref{tro} shows the Rosseland opacity from the OPAL tables \citep{1996Iglesias} as a function of temperature and density, together with the structure of the outer layers of the stellar models shown in Fig.\,\ref{env}. 
Associating an Eddington opacity with the opacity contours in Fig.\,\ref{env}, we can define the contour of $\Gamma=1$ relative to a given $L/M$. This is thus the contour where we expect to find the sonic points of our stellar models.
In Fig.\,\ref{tro} the Eddington opacities relative to our 10, 15, and 20\Msun models are plotted. They follow the increase in opacity around $\log(T)\approx 5.2$, i.e. the Fe-bump, and at $\log(T)\approx 4.6$, i.e. the HeII opacity bump. 
The more compact stellar models in Figs. \ref{env}, \ref{env10}, and \ref{env20} have their sonic points located almost exactly on their respective opacity contours and in the hot part of the iron opacity bump, i.e. $5.2<\log(T)<5.5$. 
The two more extended models showing a density inversion, i.e. the He-bump sonic point model and the  plane parallel grey atmosphere model, have their outer layers follow closely the contour of $\Gamma \approx 1$ for $ \log(T) \gtrsim 5.2$ (see Sect.\,\ref{Sect.optthickwinds}), to then cross this limit within their density inversions in the proximity of the peak temperature of the Fe-bump. The plane parallel model finds its surface back into the sub-Eddington region, while the He-bump models continues to cooler temperatures to then find its sonic point close to its Eddington opacity contour at $ \log(T) \approx 4.7$.  
The grey shaded regions in Fig.\,\ref{tro} mark instead combinations of parameters for which we do not expect to find the sonic point of radiation-driven stellar winds, due to the decrease in opacity as a function of temperature.

%The Eddington opacities corresponding to our 10 and 20 \Msun models define a region in Fig.\ref{tro} where the sonic point can be located for models with masses {\bf in the range between 10 and 20 \Msun}. Each point within the $log(\rho)-log(T)$ region {\bf in between} the contours of Eddington opacities having $ 5.2<log(T)<5.5$ corresponds to a possible $L/M$ and $\Mdot$ for which a compact model can be constructed having such sonic point conditions. %Therefore this region defines the allowed combination of density, temperature, and luminosity-to-mass ratio for the sonic point of helium stars of, in this case, galactic chemical composition, as far as their sonic point is located at $\Gamma\approx 1$. 
% 

\section{Mass-loss rate constraints}\label{sect.masslosspred}

\begin{figure}
\resizebox{\hsize}{!}{\includegraphics{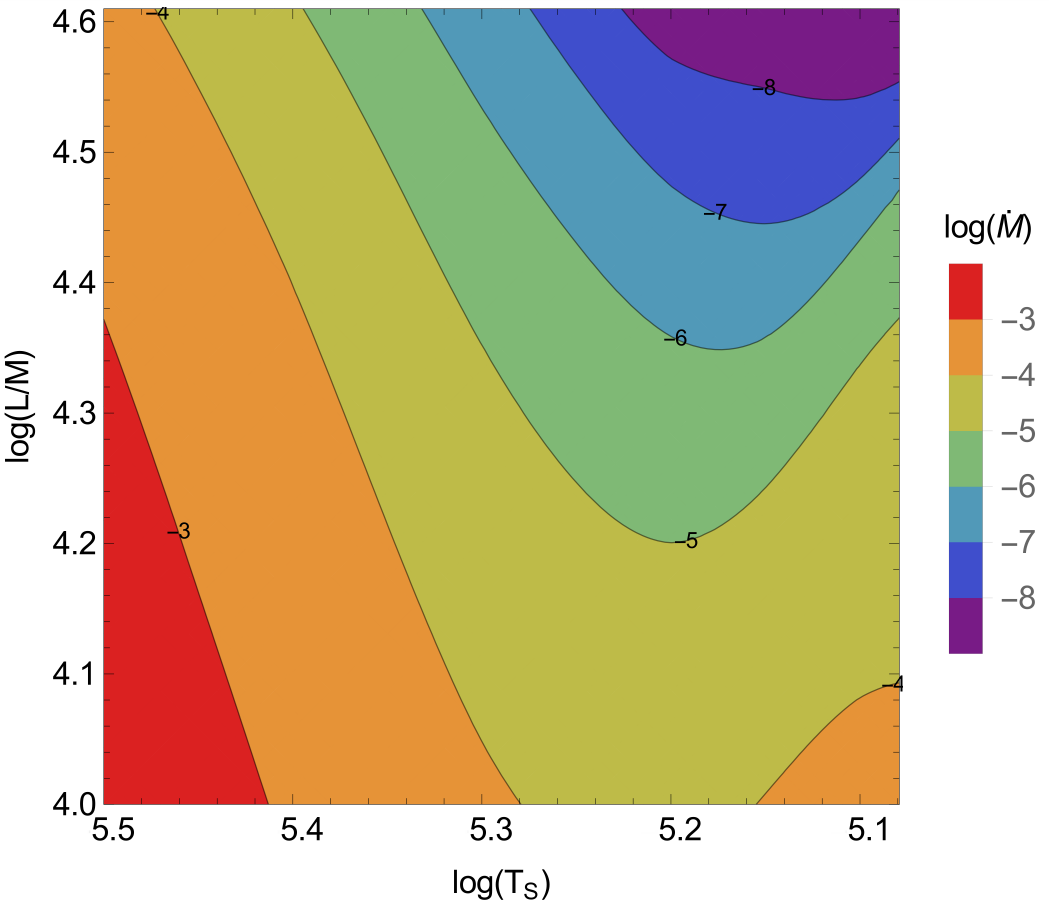}}
\caption{Contour plot showing the mass-loss rate (colour-coded  in units of $\Msun/$yr) as a function of sonic point temperature (in K) and of the luminosity-to-mass ratio (in \Lsun/\Msun) of a star. Contours are derived based on the OPAL tables for a chemical composition [{\it X,Y,Z}]=[0,0.98,0.02], with {\it X} hydrogen mass fraction, {\it Y} helium mass fraction, and {\it Z} metallicity.}
\label{mdot}
\end{figure}

\begin{figure}
\resizebox{\hsize}{!}{\includegraphics{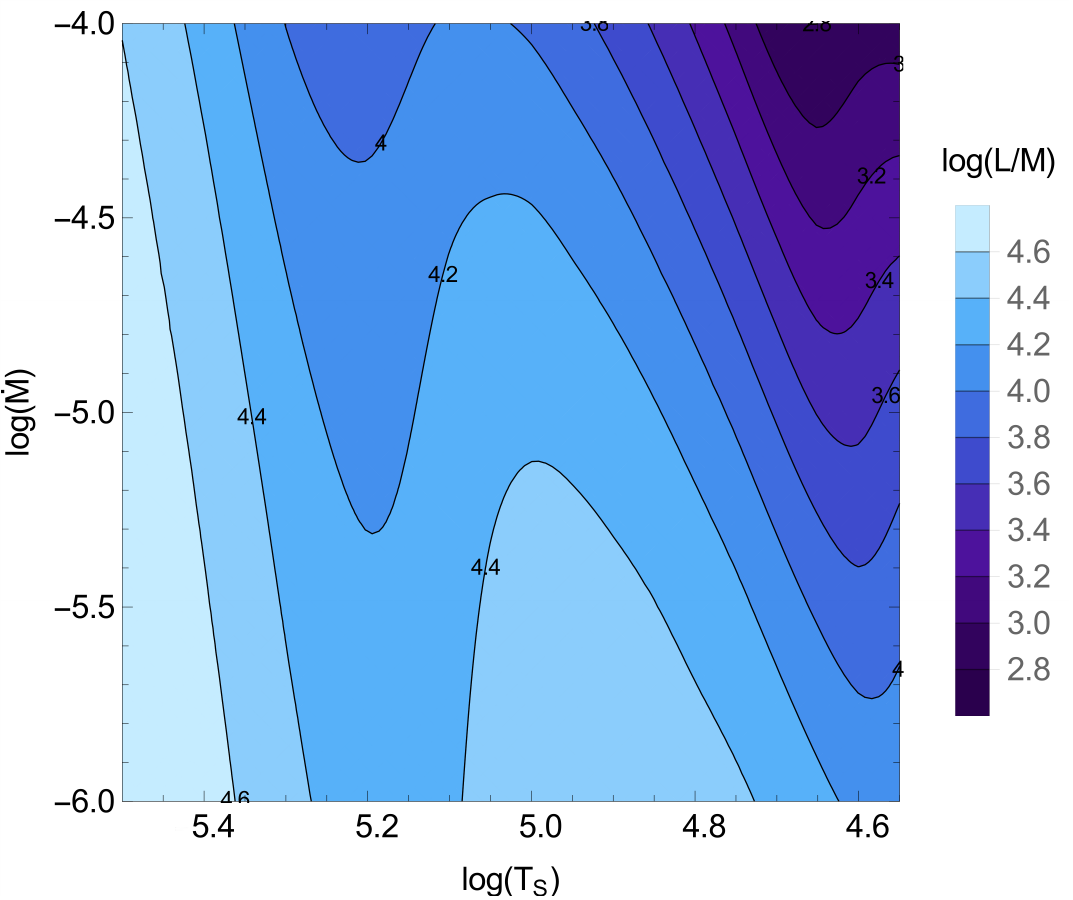}}
\caption{Contour plot showing the luminosity-to-mass ratio (colour-coded  in units of $\Lsun/\Msun$) as a function of sonic point temperature (in K) and of the mass-loss rate (in $\Msun\,{\rm yr}^{-1}$) of a star. Contours are derived based on the OPAL tables for a chemical composition [X,Y,Z]=[0,0.98,0.02], with X hydrogen mass fraction, Y helium mass fraction, and Z metallicity.}
\label{lmts}
\end{figure}

We can make use of Eq.~\ref{tsonic}, Eq.~\ref{sonicpoint}, and Eq.~\ref{kedd} to obtain the stellar wind mass-loss rates of WR stars as a function of their sonic point temperatures and luminosity-to-mass ratios. This can be done via the opacity tables, assuming as before that the value of  $\Gamma$ is exactly one at the sonic point (see Fig.\,\ref{tro} and Table~\ref{tab}). 
From this assumption it follows that the sonic point opacity is equal to the Eddington opacity, which is only proportional to the ratio $M/L$ (Eq.\,\ref{kedd}). Manipulating the OPAL tables we can unequivocally associate a sonic point density $\rho_\mathrm{ S}$ with a sonic point temperature and a given Eddington opacity (as in Fig.\,\ref{tro}). 
Combining the sonic point density and the isothermal sound speed (which is a function of the temperature only) the mass flux at the sonic point can be obtained. Once the mass flux is defined, a radius still needs to be specified to obtain a mass-loss rate (Eq.\,\ref{boundmdot}). In Fig.\,\ref{mdot} we assume a typical sonic point radius of 1 $R_{\odot}$ for all luminosity-to-mass ratios, obtaining a Sonic HR diagram\footnote{Similar to the spectroscopic HR diagram \citep{2014Langer}.} from which the mass-loss rate can be derived as a function of the sonic point temperature and the luminosity-to-mass ratio. 

Two assumptions are used to construct this diagram: the validity of the velocity independent OPAL opacities and the adoption of a specific radius to derive mass-loss rates from mass-fluxes. However, as previously discussed, the sonic point Rosseland opacities are not expected to differ significantly due to the effects of a velocity gradient, at least in the case of the Fe-bump-driven flows. Concerning the adopted radius, as shown in Sect.\,\ref{sectrisults}, solutions with the sonic point in the Fe-bump are compact, due to the lack of a strongly inflated envelope. The core and sonic point radii in the Fe-bump tend to be very similar and are not expected to differ by more than a factor of 2 from 1 $\Rsun$, at least not in the mass range of most H-free WR stars, i.e. 10--20 $\Msun$\footnote{This simplifying assumption could  be replaced by the direct computation of the sonic point radii (see our models in Sect.\,\ref{sectrisults}), but would introduce, at this point, an unnecessary complexity to Fig.\,\ref{mdot}.}. 

In Fig.\,\ref{mdot} it can be seen how the mass-loss rate contours follow the opacity profile of the iron opacity bump, as in Fig.\, \ref{tro}. Above $\log(T_\mathrm{ S})\approx5.2$, the higher  the sonic point temperature, the higher  the expected mass-loss rate at fixed $L/M$ (see  Fig.\,\ref{env}). Fixing instead the sonic point temperature, the higher the $L/M$, the lower the mass-loss rate, which  indicates that as the $L/M$ ratio increases, an Eddington factor of unity can be more easily reached, and thus flows can be accelerated to supersonic velocities already for lower mass-loss rates. Equivalently, given $\Mdot$ (thus a contour in the diagram), the flow can be accelerated until it reaches the Eddington limit, and therefore transonic velocities, already from higher temperatures for higher $L/M$ ratios. 

 In contrast, below $\log(T_\mathrm{ S})\approx5.2$ the predictions tend to be much more uncertain, mostly because sonic point temperatures lower than the iron bump peak imply that the stellar wind is not accelerated to supersonic velocities by the iron opacity bump, but rather the opacity bump leads to a low-density inflated envelope configuration with a density inversion. In this configuration the assumption of a typical radius starts to break down together with the assumption of LTE in the subsonic part of the outflow (see Table~\ref{tab}). 

Similarly, we can derive the $L/M$ ratio as a function of the sonic point temperature and the mass-loss rate.
This is done in Fig.\,\ref{lmts}, where (thanks again to Eq.\,\ref{kedd}) from the sonic point temperature and $\log(\Mdot$) we can obtain $\kappa_\mathrm{ EDD}$ and thus the $L/M$ ratio. 
For $\log(T_\mathrm{ S})\gtrsim 5.2$, higher sonic point temperatures lead to higher mass-loss rates at fixed $L/M$, or  at fixed $\Mdot$ a higher luminosity-to-mass ratio requires a higher sonic point temperature. The same trend can also be seen  in the temperature range of the raising helium opacity bump, i.e. $4.6 \lesssim \log(T_\mathrm{ S})\lesssim 5.0$, but with $L/M$ ratios which are lower at constant $\Mdot$ compared to the temperatures in the hot part of the iron opacity bump. This can be interpreted as stars needing to have higher $L/M$ in order to drive a certain $\Mdot$ while having the sonic point located in the iron bump compared to stars having their winds driven by the helium opacity bump\footnote{ WN stars cooler than $\log(T_\mathrm{ S})\lesssim 5.0$ tend to show spectra with hydrogen spectral lines, which implies non-negligible amounts of hydrogen at their surface. Thus, the H-free opacity tables might not be  adequate in that temperature range.}. Moreover, in the temperature range of the helium opacity bump, flows can be accelerated to supersonic velocities already in stars with significantly lower $L/M$ than the values necessary in the iron opacity bump.
% In fact, for example, a mass-loss rate of $\log(\Mdot)\approx -4.5$ can be accelerated to supersonic velocities by the helium opacity bump already from stars with log($L/M$)$\approx 3.6$ (see Fig.\ref{lmts}), while the same mass-loss rate would require log($L/M$)$\approx 4.5$ to achieve transonic velocities starting from the Fe-bump. 

\subsection*{Comparison to observations}

%This might be related to WNL stars, which show very high mass-loss rates while being in the HR diagram at temperatures of the order of 40\,kK \citep[see the WNL stars in ][]{2006Hamann}, or to some of the late-type WC stars \citep{2012Sander}. Indeed, WNL stars from \citet{2006Hamann} are all clustered around temperatures of the order of log$(T_*)\approx 4.6$, which may suggest that their sonic points are located in the helium opacity bump.          

Figure \ref{mdot} shows that a low-mass, low-luminosity stellar model does not have an outflow that reaches transonic velocities within the iron opacity bump unless a certain minimum mass-loss rate is applied. Lower mass-loss rates lead instead to inflated envelope configurations such as the low mass-loss rate cases in Fig.\,\ref{env} and Fig.\,\ref{env10}.

\begin{figure*}
\resizebox{\hsize}{!}{\includegraphics{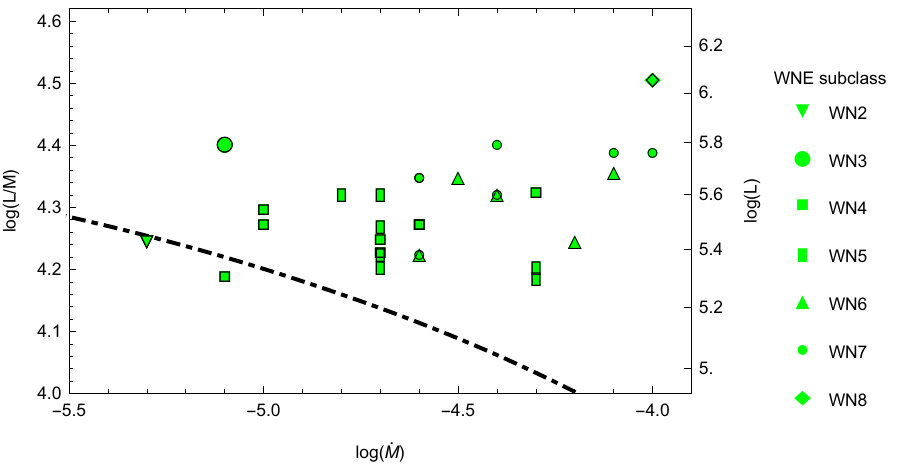}}
\caption{Minimum luminosity-to-mass ratio (in units of \Lsun/\Msun, left vertical axis) and corresponding luminosity according to the mass--luminosity relation from \citet[][right vertical axis]{1989Langer}, as a function of mass-loss rate (in $\Msun\,{\rm yr}^{-1}$) for an outflow to be driven to supersonic velocities by the radiative acceleration from the iron opacity bump. The green symbols indicate H-free WNE stars from \citet{2006Hamann}, with the different shapes indicating different spectral subclasses.}
\label{minmass}
\end{figure*}

%Only stellar models with mass-loss rates above a certain threshold have flows accelerated to supersonic velocities by the iron opacity bump, or in other words, for a certain mass-loss rate the luminosity has to exceed a threshold value for the flow to reach supersonic velocities. 
A minimum mass-loss rate as a function of the luminosity-to-mass ratio can therefore be derived, and indicates which mass-loss rate implies a radiation-driven supersonic flow starting from the iron opacity bump (minimum $\Mdot_\mathrm{ Fe}$). 
This minimum mass-loss rate  marks the separation between compact and extended solutions in Fig.\,\ref{env} and \ref{env10}. It is shown in Fig.\,\ref{minmass}, where the derived minimum $L/M$ above which the flow becomes transonic in the hot part of the iron opacity bump is plotted as a function of the mass-loss rate. The minimum $\Mdot_\mathrm{ Fe}$ decreases at higher luminosities, and it can be closely approximated by a parabola
\begin{equation}\label{eqminmasslossrate}
 \log\left(\frac{L}{M}\right) = 1.69 - 0.80\, \log\left(\Mdot\right) - 0.06\, \log\left(\Mdot\right)^2 \quad ,
\end{equation}   
or equivalently in terms of luminosity
\begin{equation}\label{eqminmasslossL}
 \log\left(L\right) = 0.44 - 1.35\, \log\left(\Mdot\right) - 0.097\, \log\left(\Mdot\right)^2 \quad .
\end{equation} 

In Fig.\,\ref{minmass} this minimum mass-loss rate as a function of $L/M$ is compared to the sample of observed Galactic WNE analysed by \citet{2006Hamann}. We collected $L/M$ and $\Mdot$ exclusively for the H-free WNE stars, as even a small mass fraction of hydrogen can significantly affect the structure of the envelopes \citep{2017Schootemeijer}. The vast majority of the observed stars exceed the minimum mass-loss rate. They are therefore consistent with models having outflows that are radiation pressure driven to supersonic velocities by the iron bump. 
%We note again that this minimum mass-loss rate has been identified just based on the assumed validity of the diffusive approximation and of the Opal opacity as a good approximation for the local flux-averaged opacities, and on the assumpion of the sonic point being at the Eddington limit. 

\begin{figure*}
\resizebox{\hsize}{!}{\includegraphics{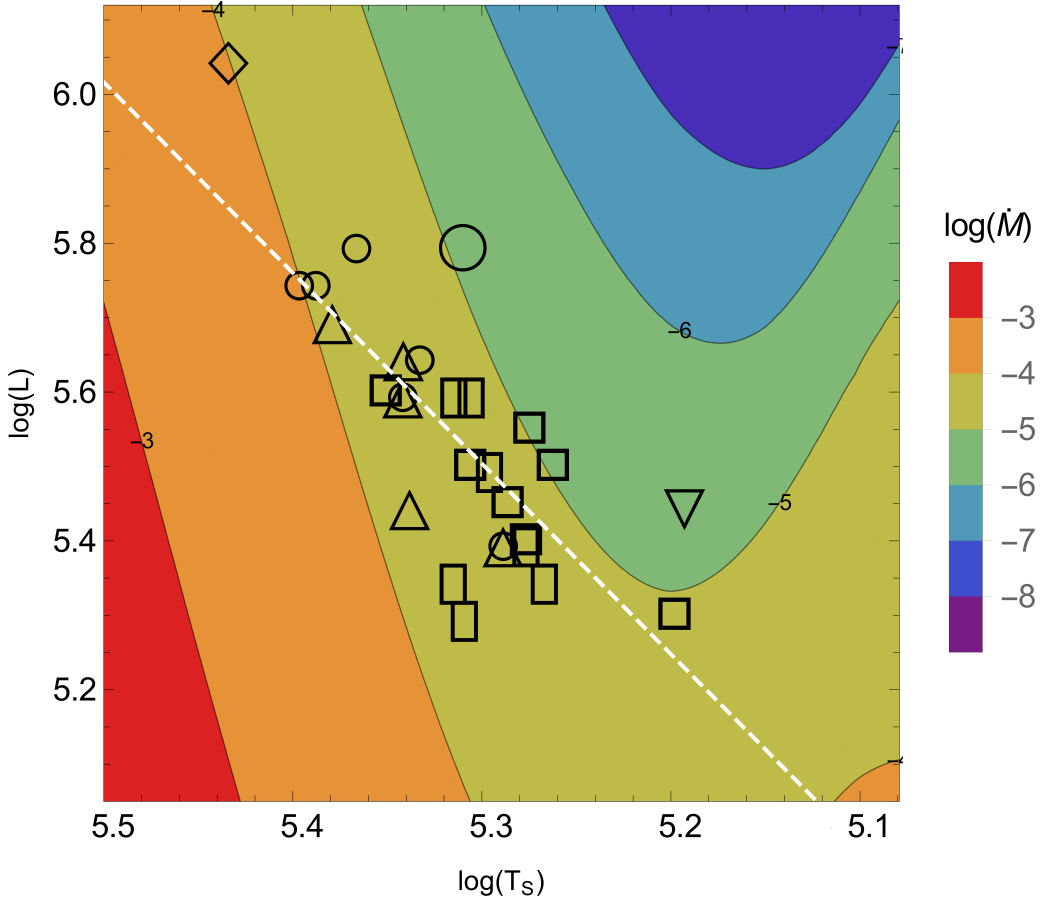}}
\caption{Sonic HR diagram relating the luminosity (in \Lsun) to the sonic point temperature (in K). The background is colour-coded according to the mass-loss rates by stellar wind (legend on the right, in units of $\Msun\,{\rm yr}^{-1}$) predicted via the use of the OPAL opacity tables and the proximity of the sonic point to the Eddington limit (see Fig.\,\ref{mdot}). The black symbols are observed Galactic WNE stars from \citet{2006Hamann}, located in this diagram via their observed luminosity and mass-loss rates and with symbols indicating their spectral subclass, as in Fig.\,\ref{minmass}. The white dashed line is the best fitting linear relation for the observed WNE stars.}
\label{mdotL}
\end{figure*}

Assuming the mass-luminosity relation from \citet{1989Langer} and, as before, a typical radius of 1\Rsun, we can construct another  Sonic HR diagram relating luminosity and sonic point temperature. This is done in Fig.\,\ref{mdotL}, where regions are colour-coded according to the expected mass-loss rate. In this diagram we locate the observed WNE stars from Fig.\,\ref{minmass}. Their location clusters at sonic point temperatures around $\log(T_\mathrm{ S})\approx 5.3$.
%For all these WR stars a compact, hot massive helium star model as the ones in Figs. \ref{env}, \ref{env10}, and \ref{env20} can be constructed, having their sonic point in the hot part of the iron opacity bump.  
Therefore, based solely on their observed luminosities and mass-loss rates, all the WNE stars of this sample can be understood as having their outflows accelerated to supersonic velocities by the radiation pressure consequence of the high number of transitions associated with iron and iron-group elements at around $\log(T_\mathrm{ S})\approx 5.2$. The sonic point temperature as a function of luminosity for the best lintear fit of the observed WNE stars in Fig.\,\ref{mdotL} follows the relation
\begin{equation}\label{eqfitLTsonic}
 \log(L)= 3.18\,\log(T_\mathrm{ S}) -11.36 \quad ,
\end{equation}
suggesting that, in general, WNE stars with higher luminosities have higher sonic point temperatures.
A similar relation as in Eq.\,\ref{eqfitLTsonic} can be derived also in terms of mass-loss rate and sonic point temperature, 
\begin{equation}\label{eqfitMDOTTsonic}
\log(\Mdot) =  5.66 \,\log(T_\mathrm{ S}) - 34.65 \quad .
\end{equation}
This time Eq.\,\ref{eqfitMDOTTsonic} suggests that higher mass-loss rates lead to higher sonic point temperatures, or in other words that stars with high mass-loss rates have their sonic point located deeper inside their atmospheres (see also Table~\ref{tab}).

%This also shows how hydrostatic stellar structure calculations are inadequate to describe the outer layers of Galactic WNE stars, as these stars have outflows that are already supersonic starting from $\approx200000$K. In fact for these objects the dynamical terms in the force balance is clearly important, and very inflated envelope solutions with $\Gamma\approx 1$ for extended regions of a star are possible only for low mass-loss rates.

\section{Discussion}

\subsection{WR radius problem}

For typical Galactic WNE stars the continuum originates in the dense outflow at a significant fraction of the terminal velocities, i.e. well beyond the hydrostatic domain \citep{2008Crowther}. Therefore, the velocity law at the base of the wind is not well constrained by observations \citep{1992Schmutz,2015Hillier}.

 As shown in Sect. \ref{sectrisults}, the subsonic structure of helium star models suggests that the location of the sonic point of WNE stars is in the range 0.9--1.5\,$\Rsun$, with supersonic flows originating from temperatures around 200\,kK (see Fig.\,\ref{mdotL}). 
%This conclusion relies only on fundamental physics arguments concerning the dynamics of the outflow and a comparison to observations. 
%As such, the sonic point radii of Galactic WNE stars seem now well constrained.
%{ Let us note here that the adoption of a typical sonic point radius to construct the Sonic HR diagram in Fig.\,\ref{mdot} and to show how observed WNE stars have sonic flows accelerated to supersonic velocities by the iron opacity bump is not leading to an internal loop  }  
 
For the supersonic layers,  \citet{2017Sander} has recently outlined how for hydrodynamically consistent models of hot, radiation-driven stellar winds the velocity profile is shaped by the different ionization levels of the elements present in the wind. 
 This suggests that the adoption of a single beta-velocity law for the supersonic wind, independently of the detailed temperature stratification, available opacities, and subsonic structure is probably an oversimplification \citep[see also ][]{2005Grafener}.
 Hydrodynamic models for WR winds might bring the hydrostatic radii in the range predicted by our stellar structure calculation \citep[see e.g. Fig.\,2 in][]{2015Sander}. A non-constant velocity gradient was found in the supersonic part of the flow of WR111, a WC5 star, for which a hydrodynamic model for the optically thick wind was available \citep{2005Grafener}. We were able to continuously connect the subsonic structure of one of our sonic point boundary conditions models and the supersonic wind model at the sonic point (see Appendix \ref{AWR111}).
 
Still, the apparent homogeneity in the location of the observed WNE stars in the Sonic HR diagram from Fig.\,\ref{mdotL} does not yet explain the scattered distribution found in the classical HR diagram \citep[cf. Fig.\,8 in][]{2006Hamann}. 
 From \citet{1989Langer}, at a given metallicity the properties of H-free post-main sequence stars are expected to be mostly defined by one single parameter, the luminosity of the star. Instead WNE stars with comparable luminosities are found to have different spectroscopic subclasses.
In this respect in Fig.\,\ref{mdotL} we note how the predicted mass-loss rate depends sensitively on the sonic point temperature. As such, small differences in the metallicity or in the subsonic force balance (e.g. due to rotation or magnetic fields), or the exact evolutionary phase and chemical composition of the core, might result in significantly different mass-loss rates, which consequently could lead to differences in the photospheric stellar parameters, which might reconcile observations with theoretical expectations.

 A decelerating flow above the sonic point could possibly lead to configurations with multiple critical points and the appearance of a density inversion similar to that observed in inflated envelopes. The presence in massive main sequence stars of an increase in density shortly after the peak temperature of the iron opacity bump has recently been confirmed via 3D simulations by \citet{2015Jiang}, implying that the density inversion is not necessarily erased, especially not by subsonic flows.
 The stagnation could then possibly imply the appearance of instabilities such as photon bubbles \citep{2003Dessart,2003Blaes,2015Owocki}, line-deshadowing instabilities \citep{1984Owocki,2013SundqvistLDI}, or stream-like lateral flows possibly connected with the co-rotating interaction regions \citep{1984Mullan,2009StLouis}. Hydrodynamic and time-dependent multi-dimensional simulations are necessary to assess the dynamical stability of such configurations and might shed light on what ultimately sets the mass-loss rate of these objects.     
  
\subsection{Optical depth of the sonic point}

Integrating the diffusive temperature gradient (Eq.~\ref{dTdr}) from the sonic point to infinity
\begin{equation}
\int_\infty^{r_\mathrm{ S}} \kappa \rho\,\mathrm{d}r = \int_{T_\infty}^{T_\mathrm{ S}}\frac{4 ac}{3 F}\,\mathrm{d}T^4 
,\end{equation}
with $a$ the radiation density constant, and assuming that the sonic point temperature $T_\mathrm{ S}^4 \gg T_\mathrm{ eff}^4 > T_\infty^4$, valid in the case of optically thick winds, leads to a tautological relation between optical depth and sonic point temperature
\begin{equation}
\tau_\mathrm{ S} \approx \frac{a c}{ 3 F} T_\mathrm{ S}^4  \quad .
\end{equation}
This, combined with Eq.~\ref{tsonic}, sets a relation which needs to be fulfilled at the sonic point for a given mass-loss rate, i.e.
\begin{equation}\label{taumdot}
\left(\frac{ 3 F}{a c}\tau_\mathrm{ S}\right)^{1/4} \approx \frac{\mu m_\mathrm{ H}}{k_\mathrm{ B}}\left(\frac{\dot{M}}{4 \pi r^2 \rho}\right)^2
\end{equation}
The mass-loss rate is a free parameter in our hydrodynamic stellar models. Equation \ref{taumdot} can then be used to constrain the only mass-loss rate which is consistent with the expected optical depth if combined with a 
prescribed velocity law for the wind \citep[as in ][]{2017Grafener}, or assuming an optical depth for the sonic point.

\subsection{Turbulence}

Atmosphere calculations of hot stellar winds often include an extra turbulent broadening of the lines, associated with the turbulent motion in the atmosphere \citep[e.g.][]{2005Dessart,2017Sander}. This turbulent velocity contributes to the equation of state of stellar matter and, via its gradients, to the structure of the outer layers\footnote{The turbulent velocity is usually assumed constant in stellar atmosphere calculation, implying that the only contribution to the force balance would arise from the density gradient.}. From the stellar structure calculations point of view, the structural effects of the inclusion of the turbulent pressure terms in the convective zones are marginal \citep{2015GrassitelliA}. Moreover, subsurface convection in the subsonic layers might be inhibited in helium star models \citep{2016Ro}. 
Therefore, we neglect the turbulent terms from our analysis, expecting at most a difference of a fraction of the gas pressure in the equation of state \citep{2015GrassitelliA,2016Grassitelli}. If present, this turbulent motion might alter our results, and introduce a systematic difference between our analysis and stellar wind models.

\subsection{Previous works}

 Compact stellar structure models were also favoured by wind models from \citet{2016Ro}. Considering the sonic point at the Fe-bump as a boundary condition for their wind models, these authors show that while using velocity independent Rosseland opacities, winds fail to accelerate up to the terminal wind velocities. However, the dynamics in the supersonic part of the flow show the presence of more compact, strong wind solutions and extended, weak stagnating solutions. The bifurcation between these solutions that  takes place depends on the ratio between the temperature scale height (computed in the diffusive limit within the supersonic part of the wind) and the local radius. In particular,  their more compact wind solutions are expected to be more prominently affected by line-force amplification due to Doppler enhancement from flow velocities of the order 150--200 km/s, suggesting that the Rosseland mean opacities become rapidly inadequate above the sonic point. The question is whether such an increase in opacity is sufficient to accelerate the outflow monotonically up to the escape speed of the star, or whether stagnation and a complex velocity profile might appear.

 After this manuscript was submitted, \citet{2018Nakauchi} constructed hydrostatic He-star models connected to a single beta-velocity law, optically thick stellar wind models. They confirm the crucial role played by the iron opacity bump in launching the winds of WNE stars, in agreement with our fully hydrodynamic models. Less satisfactory is the comparison between the observed photospheric temperature and the temperature values estimated by their wind models, further pointing towards the inadequacy of  a prescribed velocity law with a single beta exponent in reproducing the initial acceleration of these stellar winds. Similarly, \citet{2017Grafener} made use of the proximity to $\Gamma \approx 1$ at the sonic point (see Sect.\,\ref{Sect.optthickwinds}) while adopting a prescribed stellar wind dynamics to investigate WR mass-loss rates and properties. Their $\Mdot-L$ trend at Galactic metallicity resembles the observed one, favouring therefore more compact solutions with outflows driven by the iron opacity bump. Both \citet{2018Nakauchi} and \citet{2017Grafener} make use of a prescribed single beta-velocity law for the supersonic layers to constrain the mass-loss rate of WR stars with optically thick winds, while $\dot{M}$ is a free parameter in our hydrodynamic models.

In the supersonic part of WR outflows, the presence of instabilities and inhomogeneities should be taken into account. 
Invoking density inhomogeneities, or `clumping', in stellar models has an influence on the mean opacity, due to the enhanced density in clumps \citep{1988Moffat,1998Hamann}. However, a porous structure could also imply lowered mean opacity, counteracting the effect of small-scale clumping \citep{1998Shaviv,2007Oskinova}. Assuming that the material near the Fe-opacity peak is clumped, \citet{2012Grafener} were able to extend the surface radii of their hydrostatic helium star models, in practical terms, by significantly enhancing the iron bump opacity (see also Appendix \ref{app.inflation}). The surface temperatures of such hydrostatic, strongly inflated models computed with plane parallel grey atmospheres are then compared with the fictitious effective temperatures at $\tau=20$ of the atmosphere calculations performed by \citet{2006Hamann}. This effective temperature at $\tau=20$ should not be confused with the actual blanketed temperature at the base of WNE wind models, which is about a factor of 2 larger  (see e.g. Appendix \ref{AWR111}). As we show in Sect.\,\ref{sectrisults},  strongly inflated hydrodynamic solutions imply supersonic flows already at the base of the inflated envelopes for the typical mass-loss rates of WNE stars (cf. the plane parallel atmosphere model in Fig.\,\ref{env20}). Moreover, it would be inconsistent to think that the iron opacity bump is simultaneously responsible for  both the inflation of the envelope and the acceleration of the flow. 
While the detailed effects of clumping and porosity remain a subject for future research, we concentrated in this work on the homogeneous case as those instabilities are expected to initiate above the sonic point \citep{2013Sundqvist,2015Owocki}. If this is not the case and clumping is already present at the relatively high densities and optical depth of the sonic point of massive helium star models, it might affect to some extent the local opacity. 

Standing waves in helium stars \citep{1993Glatzel} are also not expected. As already pointed out by \citet{2016GrassitelliWR}, while investigating pulsations in massive helium star models, mass loss has an inhibitory effect on pulsations already in the presence of inflated envelopes.

\section{Conclusions}
%%%%%%%%%%%%%%%%%%%%%%%%%%%%%%%%%%%%%%%%%%%%%%%%%%%%%%%%%%%%%%%%%%%%%
We investigated the conditions at the sonic point and the subsonic structure of hydrodynamic models for chemically homogeneous massive helium stars, thought to be representative of WR stars in the WNE phase.
%We introduce the concept of {\it sonic horizon} for these kind of stellar models, i.e.{ :
%\begin{itemize}
% \item the subsonic part of stars with winds sufficiently optically thick becomes a zone of silence for the supersonic outflow as all perturbations in the thermodynamical quantities are advected outwards 
%\item the radiation field until the sonic point is assumed to be described by the diffusion approximation and thus based solely on local quantities, independently from the detailed conditions in the supersonic wind.
%\end{itemize} 
%For a given mass-loss rate} the sonic horizon acts like a surface that isolates the subsonic part and prevents the detailed physics of the supersonic outflow from influencing the subsonic structure. This makes WR stars the conceptual opposite of sound black holes \citep{1981Unruh,1981Thorne,1998Visser,2005Jacobson,2010Lahav}, and similar to the hydraulic white hole \citep{2011Jannes}, thought to be of utility to investigate, e.g., Hawking radiation \citep{1974Hawking}.

For typical mass-loss rates we find that the outflows of our stellar models, computed with newly implemented boundary conditions at the sonic point, are accelerated to transonic velocities by the momentum provided by the increase in opacity associated with the recombination of the iron and iron-group elements, the iron opacity bump, in the temperature range 160--220\,kK or by the helium opacity bump in the range 40--60\,kK.

The Eddington factor at the sonic point of our models is very close to unity.
This allows us to build a Sonic HR diagram relating the luminosity-to-mass ratio of stars with optically thick winds to their sonic point temperature. Knowing the sonic point temperature and relying on available opacity tables, it is possible to predict the mass-loss rates of stars with optically thick winds from this diagram.  
We use this to derive a simple relation for the minimum mass-loss rate for WNE stars necessary to have the outflow accelerated to supersonic velocities by the iron opacity bump only as a function of the luminosity-to-mass ratio.
The vast majority of the observed Galactic WNE stars exceed this minimum mass-loss rate, suggesting therefore that the radiative acceleration in the hot part of the iron opacity bump is responsible for the launch of the outflows of these stars to supersonic velocities. This leads to sonic point radii of the order of one solar radius for WNE stars.

In this paper we shed new light on the WR radius problem and  show that the iron opacity bump is key to driving outflows through the sonic point in WNE stars, and hence to determining the mass-loss rate of the star.
Our models may provide a useful reference while adopting the Sonic HR diagram to compare stellar structure and atmosphere models of optically thick outflows with observations, and may provide solid inner boundary conditions for stellar atmosphere codes, reducing the complexity of the calculations with simple physical arguments and providing ab initio subsonic structure models. 

We have also derived analytic relations to better understand the mechanisms at work when extended low-density envelopes appear in the stellar models, characterized by the proximity to the Eddington limit, the so-called envelope inflation.
In the envelope of a stellar model in hydrostatic equilibrium the emergence of a large gas pressure scale height is the response to the rapid increase in radiative opacity when approaching an opacity bump in radiation-pressure-dominated layers close to the Eddington limit.

\begin{acknowledgements} L.G. thanks Andreas Sander, Stan Owocki, Alina Istrate, Luca Fossati, and Jean-Claude Passy for the constructive discussions.  J.M. acknowledges funding from a Royal Society--Science Foundation Ireland University Research Fellowship.
\end{acknowledgements}

\bibliographystyle{aa}
\bibliography{sonic}
\appendix

\section{Envelope inflation criterion}\label{app.inflation}

\begin{figure}
\resizebox{\hsize}{!}{\includegraphics{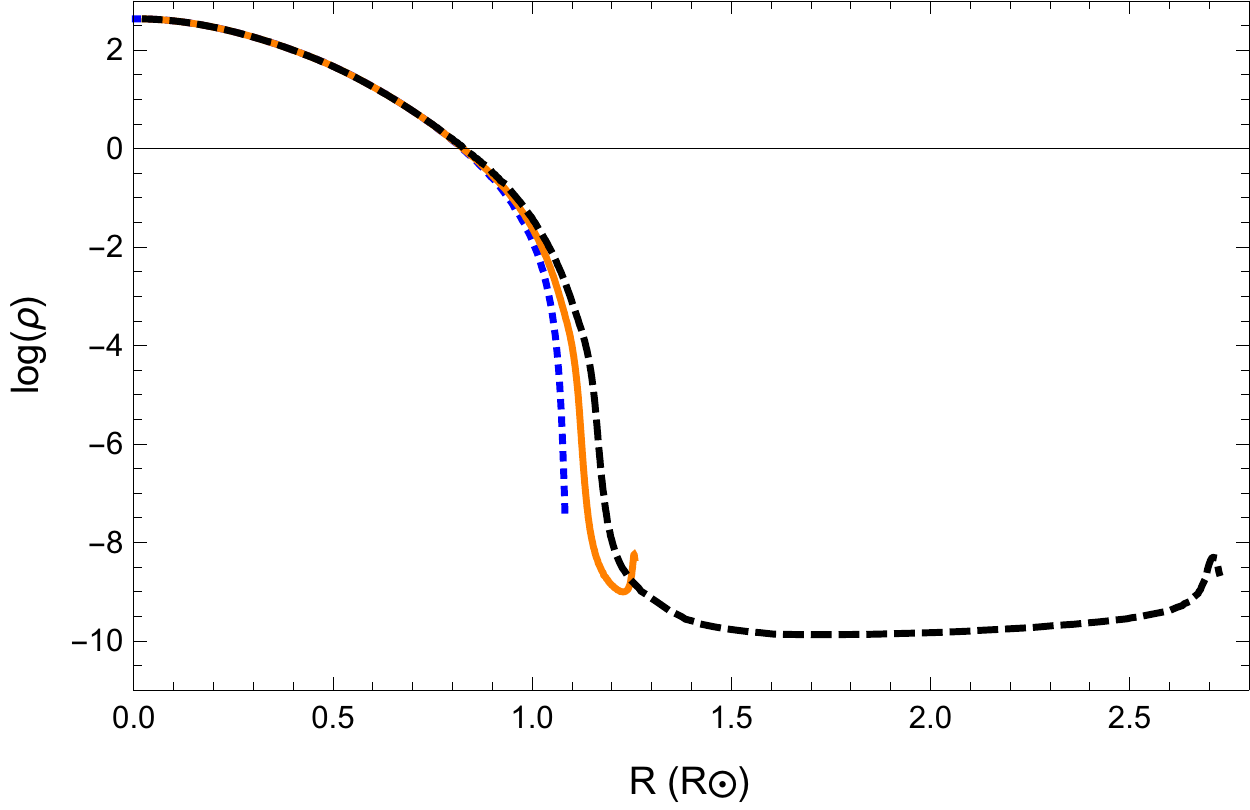}}
\caption{Density profiles in units of g/cm$^3$ of a 15$\Msun$ helium star model with different iron bump opacities. The blue dotted model is computed with the opacity table corresponding to a zero metallicity composition (i.e. without the iron opacity bump), the orange continuous model is computed with the opacity table corresponding to Z=0.02, while the black dashed model is computed with the opacity table corresponding to Z=0.05. }
\label{infpro}
\end{figure}

Here we attempt to characterize the structure of inflated envelopes. With BEC we compute three  versions of the same 15$\Msun$ helium star model, but adopting  three  different OPAL opacity tables corresponding to the opacity of stellar matter at three different metallicities. For illustrative purposes, we impose hydrostatic equilibrium and no mass loss by stellar wind. For this computation we also recover the plane parallel grey atmosphere boundary conditions \citep{1998Heger,2006Yoon}. 
The advantage of studying a helium zero age main sequence massive helium star model is that such a model is hot enough that only the iron opacity bump is present below its surface. This allows us to more clearly investigate the response of the outer layers subject only to the radiative force arising by this opacity bump.

The density profiles of these newly computed stellar models are shown in Fig.\,\ref{infpro} where the blue model is computed with the opacity table for a metallicity Z=0, while the orange and the black models are computed for Z=0.02 and Z=0.05, respectively. The Z=0 blue model clearly shows a density profile that monotonically decreases as a function of radius, showing moreover a density scale height $H_\rho$ increasingly smaller as one approaches the surface. This model shows how the density profile behaves in absence of a pronounced increase in opacity in the outer subsurface layers, showing no sign of envelope inflation and no core-halo structure below its surface. It is considered the reference model. 

This is not the case  for the helium star models with opacities corresponding to the metallicity Z=0.02 and Z=0.05. 
Their density profiles in Fig.\,\ref{infpro} show the characteristic envelope inflation configuration encountered by e.g. \citet{2012Grafener}, \citet{2015Sanyal}, and \citet{2016GrassitelliWR}. In these models the higher iron bump opacity brings the outer layers to the Eddington limit; this leads to the formation of an extended, low-density envelope which, for the Z=0.02, increases the surface radius by approximately 10\%, while for the Z=0.05 case, the increase in radius is more than a factor of 2 compared to the non-inflated case. Consequently these two models also find their surfaces at lower effective temperatures,  $T_\mathrm{eff}\approx{}$130\,kK, 120\,kK, 80\,kK for the metallicities $Z={0.00, 0.02, 0.05}$, respectively.   
At first the mildly and the more inflated models in Fig.\,\ref{infpro}  both show  density profiles with density scale heights decreasing in the core region. However, around 1$\Rsun$, the two inflated models start to differ from the non-inflated model. Their density scale heights start to increase due to the increase in the outward-directed radiative force in the proximity of the iron opacity bump.
This can be understood by writing Eq.\,\ref{densder} in the general hydrostatic case ($\varv=0$)
\begin{equation}\label{densderhydrostatic}
c^2_\mathrm{ s}\frac{\mathrm{d}\,\ln(\rho)}{\mathrm{d}r}= -g + g_\mathrm{ rad} -c^2_\mathrm{ s}\frac{\mathrm{d}\,\ln(T)}{\mathrm{d}r} \quad .
\end{equation}   
In the force balance an increase of $g_\mathrm{ rad}$ leads to a larger density scale height, and thus the high-opacity models  have more shallow density gradients. As the opacity keeps rising in the hot part of the iron opacity bump, the two  inflated models flatten their density profiles and reach a minimum, followed  by a rapid rise in density. 
%This density inversion, characterized by an increase of up to an order of magnitude of the density compared to the minimum density in the envelope, takes place according to the criterion  
From Eq.\,\ref{densderhydrostatic} (or Eq.\,\ref{densder} in the case of $\varv^2 \ll g r $ ), a positive density gradient (i.e. a density inversion) appears when 
\begin{equation}\label{gammabeta}
\Gamma \gtrsim \frac{1-\beta}{1-\frac{3}{4}\beta} 
\end{equation}
or equivalently
\begin{equation}
\frac{P_\mathrm{  gas}}{P_\mathrm{  rad}} \gtrsim 4 \frac{1-\Gamma}{\Gamma} \quad .
\end{equation}
Equaling the sides in the inequality \ref{gammabeta} gives
\begin{equation}\label{betagamma}
\beta \simeq \frac{1-\Gamma}{1-\frac{3}{4}\Gamma} \quad ,
\end{equation}
which defines the condition when either $\frac{\mathrm{d}\rho}{\mathrm{d}r}=0$ (and the flow stays subsonic) or, in the hydrodynamic case, the sonic point $\beta$ of transonic outflows (see Sect.\,\ref{Sect.optthickwinds}). Moreover, the inequality \ref{gammabeta} can be compared with the criterion for convection \citep{1997Langer}, 
\begin{equation}\label{convectioncriterion}
\Gamma \geq (1-\beta)\frac{32-24\beta}{32-24\beta-3\beta^2}
,\end{equation}
showing that a density inversion is necessarily convective \citep[see also ][]{1973Joss}.

Equation \ref{betagamma} shows that a null density gradient can appear only for specific sets of $\beta$ and $\Gamma$, with lower $\beta$ (and thus densities) for higher $\Gamma$, as we can see in Fig.\,\ref{infpro}. A modified Eddington factor could then be introduced, namely
\begin{equation}\label{gammagrassitelli}
\Gamma_\mathrm{ S}=\frac{\kappa L}{4 \pi c G M}\left(1+\frac{P_\mathrm{  gas}}{4 P_\mathrm{  rad}}\right)
,\end{equation}
which includes the outward-directed force associated with the gas temperature gradient, and which defines (when equal to unity) either the beginning of a density inversion or the condition at the sonic point of stars for which convection is inefficient (see Sect.\,\ref{Sect.optthickwinds} and Eq.\,\ref{gammasonic}). 
This equation also shows that the local Eddington factor cannot be exactly unity either at the sonic point or when $\frac{\mathrm{d}\rho}{\mathrm{d}r}=0$. In either case, these two quantities tend to differ by only few percent in the radiation-pressure-dominated  outer layers of massive stars, which is in fact equivalent to the approximation $\frac{\mathrm{d}c_\mathrm{ s}^2}{\mathrm{d}r} - 2 \frac{c_\mathrm{ s}^2}{r} \ll g$ made in Sect.\,\ref{Sect.optthickwinds} and below. 

From Fig.\,\ref{infpro} it is evident that the inflated stellar models differ markedly from the non-inflated model when the profiles show an inflection point, i.e. a minimum in $\mathrm{d}\,\ln(P_\mathrm{  gas})/\mathrm{d}r$, or almost equivalently $\mathrm{d}\,\ln(\rho)/\mathrm{d}r$. An increase in the density scale height follows the inflection point, contrary to the monotonically decreasing density scale height of stellar models which do not inflate. 
This is due to an increasing contribution from $g_\mathrm{ rad}$ in the momentum equation, affecting the balance between the gravitational force and the gas pressure gradient that defines the condition of hydrostatic equilibrium. 
Analytically, the second derivative of the gas pressure profile writes
\begin{equation}\label{seconder}
\begin{aligned}
\frac{\mathrm{d}^2\ln(P_\mathrm{  gas})}{\mathrm{d}r^2}={}&\frac{\mathrm{d}^2\ln(\rho)}{\mathrm{d}r^2}+\frac{\mathrm{d}^2\ln(T)}{\mathrm{d}r^2}  \\  {}&= \frac{(g-g_\mathrm{ rad})}{c_\mathrm{ s}^2}\frac{\mathrm{d}\,\ln(T)}{\mathrm{d}r} +\frac{g_\mathrm{ rad}}{c_\mathrm{ s}^2} \frac{\mathrm{d}\,\ln(\kappa)}{\mathrm{d}r}
\end{aligned}
,\end{equation}   
where for simplicity we have neglected the geometrical term as $2/r \ll \mathrm{d}{\rm ln}(T)/\mathrm{d}r$ (which might not be always possible for main sequence stars). In writing Eq.\,\ref{seconder} we have assumed a fixed chemical composition and no energy source or sink in the outer layers (strictly true for hydrostatic models in thermal equilibrium, and true to a high degree for steady-state hydrodynamic models).

In Eq.\,\ref{seconder} the first term on the right-hand side is always negative in sub-Eddington regions. Instead, the second term on the right-hand side depends on both the radiative acceleration and the opacity profile, and it is positive when the opacity increases with increasing radius. This is the term that becomes dominant at and above the inflection point. For a constant opacity, an inflection point would be present only when $\Gamma = 1$. However,  in the case of more complex, non-constant opacity profiles, a rapid increase in opacity at temperatures where the recombination of some elements takes place can lead to an increase in the gas pressure scale height, and thus density scale height, as the second derivative of gas pressure changes sign. 

From Eq.\,\ref{seconder} a criterion for the appearance of the inflection point, i.e. for the beginning of an envelope inflation can be derived:
\begin{equation}\label{critinf}
\frac{\mathrm{d}\,\rm{ln}(\kappa)}{\mathrm{d}\rm{\, ln}(T)} \equiv \frac{4 \, \mathrm{d}\rm{\, ln}(\Gamma)}{\mathrm{d}\rm{\, ln}(P_\mathrm{  rad})} \lesssim 1 -\frac{1}{\Gamma}
\end{equation}
We can now understand envelope inflation as {the appearance of low gas pressure gradients in response to a steep increase in the radiative opacity in radiation-pressure-dominated layers close to the Eddington limit}. Above the inflection point, the increase in opacity increases the gas pressure scale height $H_{P_\mathrm{  gas}}$ following the increase in $g_\mathrm{ rad}$ as $\Gamma \rightarrow 1$ \citep{2017Sanyal}: 
 \begin{equation}\label{gasscaleheight}
\frac{1}{H_{P_\mathrm{  gas}}} := \frac{\mathrm{d}\rm{ln}(P_\mathrm{  gas})}{\mathrm{d}r}=\frac{g-g_\mathrm{ rad}}{c_\mathrm{s}}= \frac{g}{c_\mathrm{s}} (1-\Gamma)  .
\end{equation}   
 From here it follows that as the gas pressure scale height increases, the location of the photosphere is found at larger radii. Analytical expressions for the latter in the case of envelope inflation originating from the iron bump in very massive helium stars and neglecting convection have been given in \citet{2012Grafener}.
 We note here that stellar models can be inflated, i.e. have extended low-density envelopes close to the Eddington limit, but do not show a density inversion \citep[or a gas pressure inversion; see e.g. the 85$\Msun$ main sequence model in ][]{2015Sanyal}. On the other hand, a density inversion appearing when $\Gamma\approx 1$ is always an indication of envelope inflation.

\subsection*{Effects of convection on inflated envelopes}
The presence of convection can affect the structure of the inflated envelope \citep{2015Sanyal}. It effectively reduces the radial extent of the inflated layers by affecting the temperature stratification without contributing directly to the force balance. In convective layers, the ratio of the luminosity transported by radiation to the total luminosity can be expressed as 
\begin{equation}\label{nablaL}
\frac{L_\mathrm{ rad}}{L}=\frac{\nabla}{\nabla_\mathrm{ rad}}
,\end{equation}
where $\nabla_\mathrm{ rad}$ is the temperature gradient required to transport all the luminosity by radiation, while $\nabla$ is the actual temperature gradient which includes the contribution from convection. In a convective layer, according to the Schwartzschild criterion, $\nabla_\mathrm{ rad}>\nabla$ \citep{1990Kippenhahn}. Together with Eq.\,\ref{nablaL}, the radiative acceleration from Eq.\,\ref{grad} writes as
\begin{equation}
g^{\rm conv}_\mathrm{ rad} = \frac{\kappa L }{4 \pi r^2 c}\frac{\nabla}{\nabla_\mathrm{ rad}} < g_\mathrm{ rad} \quad ,
\end{equation}
where the inequality arises from $\nabla_\mathrm{ rad}>\nabla$. 
Therefore, convection effectively reduces the radiative acceleration by not contributing to the force balance or, equivalently, has the same effect of a reduction of the local opacity as 
\begin{equation}
\kappa_\mathrm{ conv}=\frac{\nabla}{\nabla_\mathrm{ rad}} \kappa \quad,
\end{equation}
leading to small gas pressure scale height (Eq.\,\ref{gasscaleheight}). The criterion in Eq.\,\ref{critinf} is still valid, as far as the opacity gradient takes into account the effect of convection.

\section{Radiation field at the sonic point}\label{SH}

In general, the temperature stratification in a stellar atmosphere is a global problem due to the intrinsic coupling of the radiation field to different regions of the atmosphere. However, we  discuss here the limiting case in which the radiative energy transport turns from being a global problem, to a local one, and apply it to our analysis of the conditions at the sonic point of WNE stars. 
%By no means the following considerations should be taken as guaranteed for all WR stars, but should rather be tested via detailed atmosphere calculations in different contexts. 

The intensity of radiation $I$ at the sonic point optical depth $\tau_\mathrm{ S}>1$ in the Rosseland approximation can be expressed in terms of a Taylor--McLaurin series around the equilibrium Planckian value $B(T)$ of the source function. 
The source function $S$ can be written as \citep{1978Mihalas,1994Hansen}
\begin{equation}\label{sourcefunct}
S(t)=\sum\limits_{n=0}^\infty \frac{(t-\tau_\mathrm{ S})^n}{n!} \frac{\mathrm{d}^n B(\tau_\mathrm{ S})}{\mathrm{d}\tau^n} \quad ,
\end{equation}
with $B(\tau_\mathrm{ S}):=B(T[\tau_\mathrm{ S}])$ the Planck function at the temperature of the sonic point and $t$ a dummy variable.\\
Integration of the equation of radiative transfer
\begin{equation}
\frac{\partial I e^{-\tau/\mu}}{\partial \tau}=-\frac{1}{\mu} S(t) e^{-\tau/\mu}
\end{equation}
for the outgoing radiation $I^+$, i.e. from $\tau=\infty$ to $\tau=\tau_\mathrm{ S}$, leads to
\begin{equation}\label{radtransfequation+}
I^+(\tau_\mathrm{ S}, \mu\geq 0) = \sum\limits_{n=0}^\infty \mu^n \frac{\mathrm{d}^n B(\tau_\mathrm{ S})}{\mathrm{d}\tau^n}
,\end{equation}
with $\mu:=cos(\theta)$ indicating the direction of the beam. 
The expression for the incoming radiation $I^-$ from $\tau=0$ to $\tau=\tau_\mathrm{ S}$ writes instead 
\begin{equation}\label{radtransfequation-}
I^-(\tau_\mathrm{ S}, \mu\leq 0) = \sum\limits_{n=0}^\infty \mu^n \frac{\mathrm{d}^n B(\tau_\mathrm{ S})}{\mathrm{d}\tau^n}\left( 1 - e^\frac{\tau_\mathrm{ S}}{\mu} \sum\limits_{k=0}^n \left(\frac{-\tau_\mathrm{ S}}{\mu}\right)^k\frac{1}{k!} \right) \quad.
\end{equation}
The  inward-directed radiation differs from the outward-directed radiation by terms of the order of $e^\frac{\tau}{\mu}$.

We can now more quantitatively refer to WNE stars. At first we can say that, assuming as reference a typical optical depth for the sonic point of the order of $\tau_\mathrm{ S} \approx 5$ \citep[see e.g. Table~\ref{tab} and][]{2005Grafener}, the exponential term in Eq.\, \ref{radtransfequation-} contributes for less than 1$\%$. For low $n$ therefore, the term $e^\frac{\tau}{\mu} \sum\limits_{k=0}^n \left(\frac{-\tau}{\mu}\right)^k\frac{1}{k!}$ is negligible, as usually assumed in order to derive the diffusive approximation \citep{1978Mihalas}. 
Considering the momenta of the radiative energy transport equation, namely the energy density, the energy flux, and the radiation pressure, an approximation of the derivatives by appropriate differences shows that the ratio of successive terms in the series is of order O(1/$\tau_\mathrm{ S}^2$) (see \citealt{1978Mihalas}). As such, we can study the convergence of the series via
\begin{equation}
O\left(\frac{1}{\tau_\mathrm{ S}^2}\right)\approx O\left(\frac{\lambda^2}{H_\mathrm{ P}^2}\right) \quad ,
\end{equation}
where we have taken the pressure scale height $H_\mathrm{ P}$ as the characteristic scale of the system. From Table~\ref{tab} the pressure scale heights are an order of magnitude larger than the mean free path at the sonic point, assuring the rapid convergence of the series to the Planckian value, with the high-order terms contributing less than 1$\%$.  
We can thus assert that the conditions at the sonic point of WNE stars, especially those with stellar winds driven by the iron opacity bump, closely approach the LTE conditions. 

Consequently, the radiation field at the sonic point of these stars is close to being isotropic, with the radiative transfer losing the explicit dependence on the optical depth in Eq.~\ref{radtransfequation-}. We can further estimate the level of anisotropy of the radiation field at the sonic point in case of outflows driven by the iron opacity bump as \citep{1978Mihalas}
\begin{equation}
\frac{\textrm{Anisotropic term}}{\textrm{Isotropic term}}\propto\left(\frac{T_\mathrm{ eff}|_\mathrm{ S}}{T_\mathrm{ S}}\right)^4\approx 5\%
,\end{equation}
with the $T_\mathrm{ eff}|_\mathrm{ S}$ fictitious effective temperature at the sonic point derived from the Stefan--Boltzmann law, having adopted a radius of 1$\Rsun$ and a typical luminosity of $10^5 \Lsun$. 
We can also estimate the proximity to the LTE conditions and the validity of the diffusive approximation by comparing the radiation timescale associated with the photon thermalization
\begin{equation}
\tau_{t}=\frac{\lambda}{c}
\end{equation}
to the other macroscopic timescales, i.e. the dynamical and thermal timescales of the envelopes \citep{1981Thorne,1981ThorneB}. 
The timescale for photon thermalization, i.e. the time needed by the radiation field to approach its LTE value, is of the order of seconds, while the dynamical and thermal timescales were estimated to be of the order of hundreds of seconds for massive helium stars \citep{2016GrassitelliWR}. 

All these considerations lead us to conclude that the states of the gas and radiation are close to their LTE conditions.
In LTE, it is assumed that the local T and $\rho$ are sufficient to determine the ionization state of the gas,  all microscopic processes being close to detailed balance. 
%The assumption of LTE has a crucial implication: it makes the radiative transfer problem a {\it local} problem.  
This approximation is justified as far as the thermalization depth of radiation, i.e. the depth at which the source function approaches its equilibrium value, is smaller than the sonic point optical depth \citep{1978Mihalas,1995Pistinner}.

%A crucial physical consideration to be done is that absorption and emission processes are intimately linked to the local thermodynamical quantities, while scattering of photons tend to avoid strict coupling with the local quantities, being instead controlled by the global properties of the stellar atmosphere \citep{1984Mihalas}. When the coupling of the line-emission process to the thermal pool happens to be weak, the source function can depart drastically from the LTE values, which implies that non-local effects can dominate the form of the source function \citep{1978Mihalas}. Moreover, of course, the frequency dependence of the radiation field makes the problem more complex. When instead the collision rate is greater than the radiative rates, one can essentially recover the LTE conditions.  

\section{Nozzle analogy and critical point}\label{appendixsoniccritical}

%In order to understand the physical implications arising in the case of optically thick stellar winds we can draw a parallel with the physics that took mankind to the Moon. In fact 
The {critical solution} describing the kinematics of stellar winds
%, namely the only solution of the momentum equation having everywhere a positive velocity gradient, 
can be understood in analogy with an ideal rocket (or de Laval) nozzle, i.e. a nozzle with walls that narrow at first, reach a throat, and then rapidly expand \citep{1980Abbott,1995Bjorkman,1999LamersCassinelli,2007Shore}. Given a compressible steady-state flow of gas,  the subsonic flow accelerates in the converging part of the nozzle due to mass conservation. If the initial conditions are such that the flow can reach supersonic velocities before passing the throat, the sonic point is always located exactly at the throat of the nozzle.
%The regulation mechanism for the adjustment of the flow relies on sound-waves (or in general to the propagation of information) traveling upstream and affecting the thermodynamic quantities in the subsonic part of the nozzle. 
This system, common to almost all modern rockets \citep{1989Raga}, shows how a smooth steady-state flow can self-adjust such that a transonic flow can be established via information that travels at the speed of sound. Once the transonic flow is established, the conditions in the diverging supersonic part define the local kinematics of the flow,  such as its velocity and temperature, but they can influence neither the subsonic part of the flow nor the mass-flow rate \citep{1999LamersCassinelli,2007Shore,2014Maciel}. 

The analogy with a stellar wind lies in the form of the momentum equation of a de Laval nozzle, that is \citep{1999LamersCassinelli}
\begin{equation}
\varv^2 \frac{\mathrm{d}{\rm ln} \varv}{\mathrm{d}x} =  -c_\mathrm{ s}^2\frac{\mathrm{d}{\rm ln} \rho}{\mathrm{d}x} -\frac{\mathrm{d} c_\mathrm{ s}^2}{\mathrm{d}x}  
\end{equation}
for an ideal gas flow, with $x$ spatial coordinate. Combining this with the continuity equation $\mathrm{d}{\rm ln}(\varv) + \mathrm{d}{\rm ln} (\rho) + \mathrm{d}{\rm ln} (S)=0$, where S corresponds to the area of a section of the nozzle, the force balance is given by
\begin{equation}\label{nozzleeq}
\left(\varv^2 - c_\mathrm{ s}^2\right)\frac{\mathrm{d}{\rm ln} \varv}{\mathrm{d}x} =  -c_\mathrm{ s}^2\frac{\mathrm{d}{\rm ln} S }{\mathrm{d}x} -\frac{d c_\mathrm{ s}^2}{\mathrm{d}x}  \quad .
\end{equation}
Comparing Eq.\,\ref{nozzleeq} with Eq.\,\ref{criticalpoint}, the pseudo-nozzle term in the context of radiation-pressure-driven winds is \begin{equation}
c_\mathrm{ s}^2 \frac{\mathrm{d}{\rm ln} S }{\mathrm{d}x} = g \left(1-\Gamma + \frac{2 c_\mathrm{ s}^2}{g r}\right)
,\end{equation} 
which shows  how, in the limit of $2c_\mathrm{ s}^2\ll gr$, the Eddington limit acts similarly to a nozzle term for a radiation-driven stellar outflow.

The common property between the hydrodynamic of a flow through a pipe and a stellar wind is the readjustment to the characteristic solution. It is the solution that starts subsonic, goes through the critical point of the momentum equation \citep[where the mass-flow is defined,][]{1999LamersCassinelli}, and has a finite velocity at infinity \citep{1980Abbott,2002Nugis,2007Shore}. 
 In the nozzle the critical point also defines the last point of the system which can still communicate with all the regions of the flow. 
 By reducing the perturbative analysis in e.g. \citet{1980Abbott} to a case where the line opacity and acceleration are independent from the velocity profile, we can easily show that the characteristic speed of the hyperbolic system of stellar wind equations is the local sound speed\footnote{\citet{1980Abbott} introduced the concept of effective sound speed and radiative-acoustic waves while treating the CAK critical point.}.  However, as noticed by \citet{1981Thorne} while investigating the event horizon, the diffusive approximation for radiative transfer is notoriously acausal (due to the parabolic nature of the partial differential equation governing the phenomenon). This shortcoming implies that information can be transmitted instantaneously. Hence a relativistic formulation governing diffusion together with the introduction of the  second sound speed\footnote{The {second sound speed} is the speed at which a perturbation in the heat flux travels, i.e. the speed of heat. It arises when the Fourier equation for diffusive transport gets a hyperbolic form via a relativistic description of the phenomena \citep{2005Ali} } would be more appropriate \citep{1982Flammang,1984Mihalas,2001Mandelis,2005Ali}.  If the speed of heat were, in this context, the same as or smaller than the speed of sound, a {sonic horizon} would arise in this class of stars, with the sonic point being the last point that could communicate with all the regions of the flow.

%In general terms, a so called critical point occurs whenever the Jacobian matrix associated with the set of differential equations describing a system has vanishing elements, and, as divergent derivatives have no physical significance, extra conditions such as regularity conditions have to be specified \citep{1988Nobili}. The sonic point is a well know example of a critical point in the astrophysical context \citep[e.g. ][]{1966Parker,1980Abbott,1981ThorneB,1984Matsumoto,1995Bjorkman}. 
\subsection*{Effects of velocity gradients on the critical point}

Equation \ref{criticalpoint} shows that the sonic point is the critical point of a stellar wind, but this is true only if no other term with a dependency on $\mathrm{d}\varv/\mathrm{d}r$ contributes to the right-hand side of Eq.\,\ref{criticalpoint}. In the context of line-driven winds, \citet{1975CAK} and later applications of their theory (the CAK theory) showed the importance of lines unsaturated due to Doppler shift in driving the winds of massive stars. In fact, in the presence of steep velocity gradients the rest wavelength of each accelerating element in the outflow is Doppler shifted such that unattenuated continuum can be absorbed and momentum transfer is very efficient. This in turn allows massive stars to have winds with very high terminal wind velocities. In this context the opacity arising from the spectral lines includes a dependency on the velocity gradient,  i.e. $\kappa(\rho,T,\mathrm{d}\varv/\mathrm{d}r,...)$, and therefore \citet{1980Abbott} argue that the sonic point is no longer, strictly speaking, the critical point of the outflow. 

However, \citet{2002Nugis} suggest that the contribution to the opacity by the Doppler shifted lines in hot WR stars is usually less than 1\% at typical sonic point conditions. This can be quantified by investigating the inverse of the optical depth parameter  (Table~\ref{tab}). This quantity compares the Doppler shift of the lines as being due to the velocity gradient through a mean free path to the thermal velocity of the protons \citep{2002Nugis,2016Ro}. They estimate this quantity to be typically of the order of 100 for WR stars.  %We test this hypothesis via our stellar structure calculations (Sect.\,\ref{sectrisults}), testing therefore whether such extra contribution from Doppler shifted lines in the computation of the opacity could be neglected, and whether the adoption of the velocity independent Rosseland opacities from the standard OPAL tables for non-expanding media is reasonable until the sonic point. 
Independently of this, \citet{2007LucyB,2007Lucy} also suggested that, in contrast to the standard CAK {\it ansatz}, the sonic point retains its role of critical point for the system of differential equations describing the system \citep[see also ][]{1975Lucy,1998Lucy,1990Poe,2008Muller,2017Sander}. The importance of the unsaturated lines in the highly supersonic outflow is instead clear \citep{1975CAK,1980Abbott,1985Abbott,1989Kudritzki,1995Gayley,2005Grafener,2005Vink,2008Puls,2008Grafener}.

\section{WR111}\label{AWR111}

\begin{figure}
\resizebox{\hsize}{!}{\includegraphics{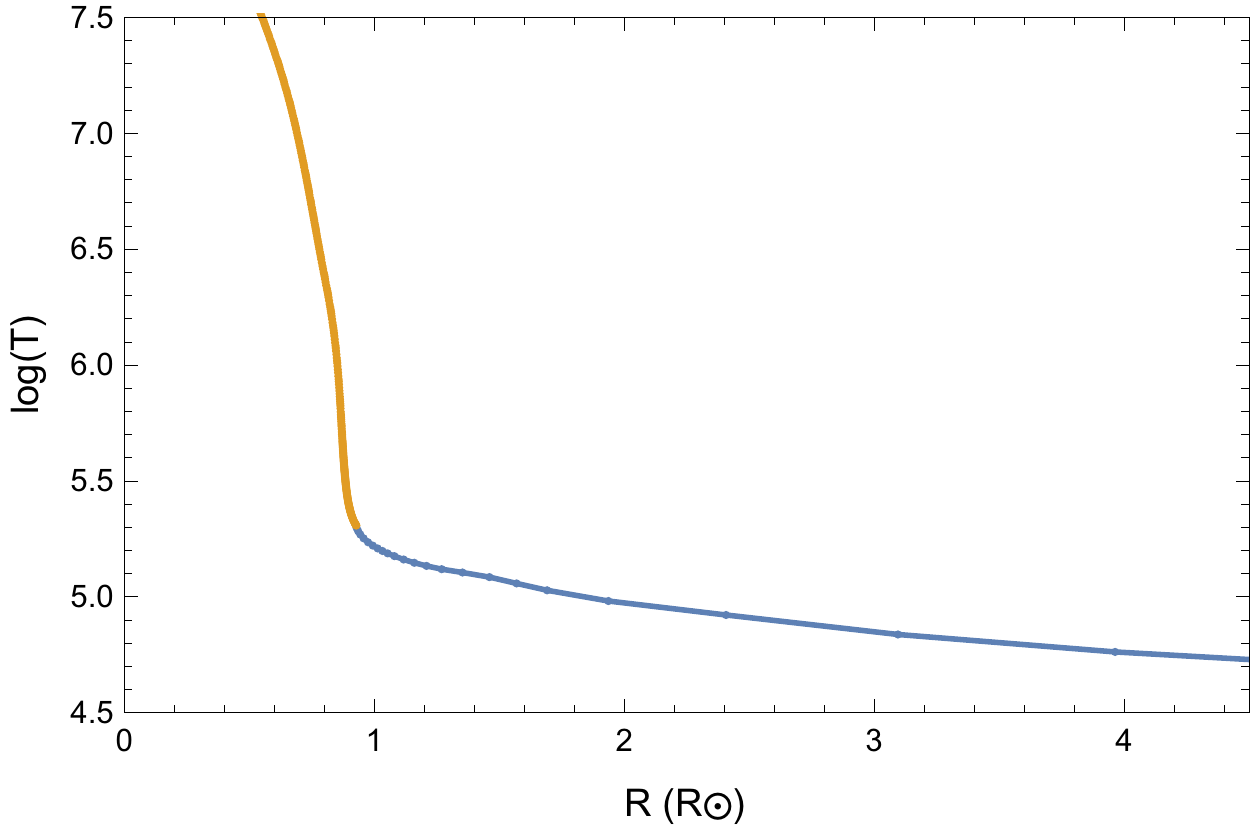}}
\resizebox{\hsize}{!}{\includegraphics{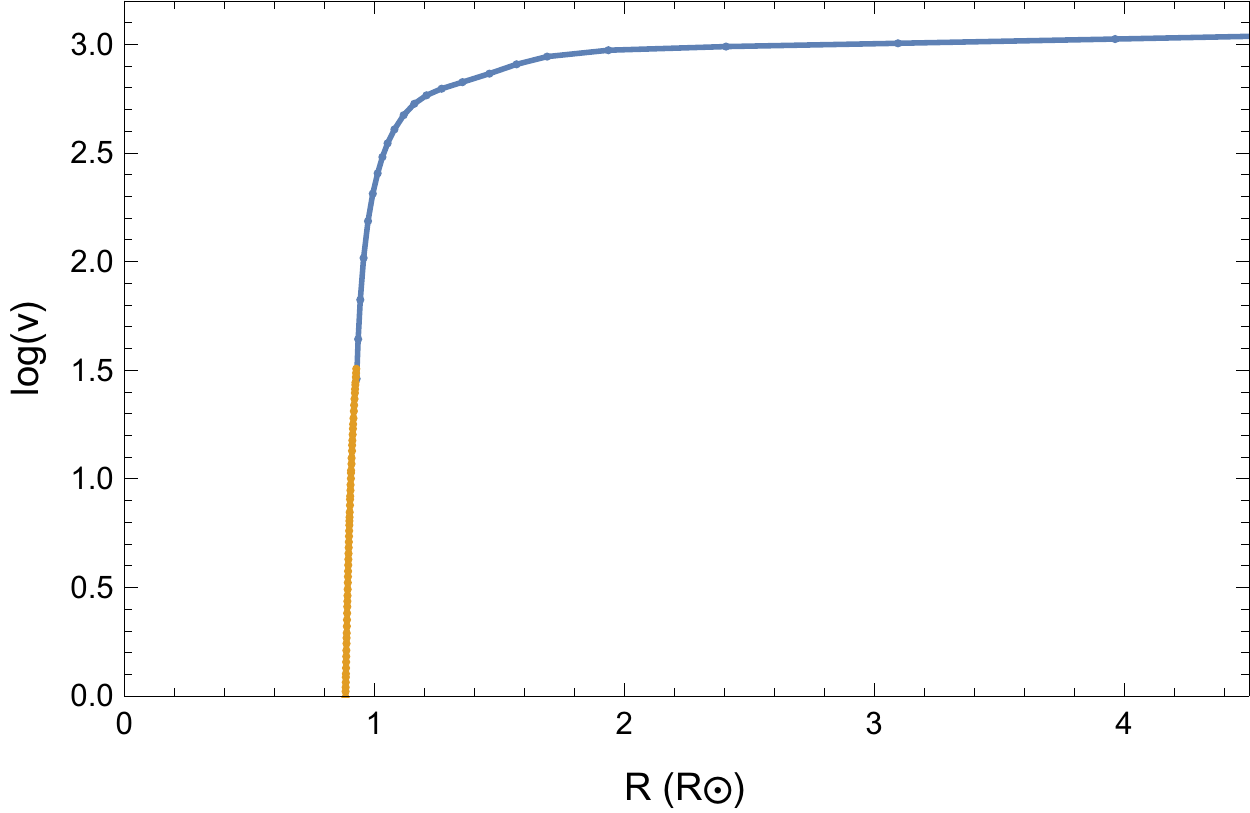}}
\caption{ Temperature (in K) and velocity profile (in km/s) as a function of radius for the WR111 stellar structure and wind model. The orange profile indicates the subsonic structure derived with our newly developed models, while the blue profile is the supersonic wind profile from \citet{2005Grafener}.}
\label{WR111}
\end{figure}

We adopt the hydrodynamically self-consistent wind structure derived for WR111, a WC5 WR star, by \citet{2005Grafener} and connect it smoothly to our hydrodynamic model. This is done in order to  show how our calculations of the subsonic structure can be smoothly matched to the sonic point conditions of a hydrodynamical wind model of a post-main sequence WR star and also to give a picture of what the internal profile of such a star looks like. This is done in Fig.\,\ref{WR111}, where the velocity and temperature profiles of a 12$\Msun$ stellar model with sonic point boundary conditions are plotted together with the supersonic structure of the wind \citep{2005Grafener}. The stellar model has been built such that its sonic point conditions, namely sonic point temperature, density, radius, luminosity, and mass-loss rate, match those derived for WR111.
%This gives a picture of the attempt to obtain a fully consistent model of the internal structure of a WR star by combining stellar structure and wind calculations. 
Luminosity and mass loss for the stellar model are, following \citet{2005Grafener}, $\log(L/\Lsun)=5.45$ and $\log(\Mdot)\approx -5.1$, respectively.  
Homogeneous zero age main sequence helium star models with such luminosity were found to have a  core radius that was too large to reproduce the radius of this wind model. Therefore, given that WR111 is of the WC class and it is consequently thought to be in a more advanced stage of its evolution, we used a more evolved model (with approximately 0.5 carbon mass fraction in its centre) instead of a homogeneous helium star. 
%This is also helpful in showing how compact models, having their radii mostly defined by the hydrostatic radii of the core, have consequently sonic radii which can not vary considerably (i.e. by more than a factor unity). Inflation alone can not be held responsible for the WR radii problem, as the iron opacity bump is necessary to accelerate the flow to transonic velocities \citep[see Sect.\,\ref{sect.masslosspred} of this manuscript and][]{2008Crowther}. 
The wind model by \citet{2005Grafener} found its sonic point at an optical depth of $\tau=5.4$ with temperature $T_\mathrm{ S}\approx 200\,kK$ and $\log(\rho_\mathrm{ S})\approx 8.5$. For this star \citet{2005Grafener} finds a $T_*$, i.e. the effective temperature at $\tau=20$, equal to 140\,kK. Instead, for the same object, spectral analysis conducted by \citet{2002Grafener} and \citet{2012Sander} assuming a beta-velocity law led to $T_*$ of the order of 85\,kK. The corresponding radius derived via the Stefan--Boltzmann law would thus be $\approx 2.5 \Rsun$, different by more than a factor of 2 from the actual quasi-hydrostatic radius in Fig.\,\ref{WR111}. This shows  how misleading it might be to use  the Stefan--Boltzmann law at $\tau=20$ and, more importantly, how the adoption of a beta-velocity law to describe the dynamics of the optically thick winds of WR stars might lead to a significantly different radius estimate compared to hydrodynamic calculations.

\end{document}